\documentclass{statsoc}

\usepackage{caption}
\usepackage{subcaption}

\usepackage[a4paper]{geometry}
\usepackage{graphicx}
\usepackage[textwidth=8em,textsize=small]{todonotes}
\usepackage{amsmath}
\usepackage{amsfonts}
\usepackage{natbib}
\usepackage{float}
\restylefloat{table}
\allowdisplaybreaks

\usepackage{todonotes}

\newcommand{\noi}{\noindent}

\newcommand{\laspace}{\;}
\newcommand{\mespace}{\:}

\newcommand{\intspace}{\mespace}
\newcommand{\isp}{\intspace}

\newcommand{\eg}{{\em e.g.\ }}
\newcommand{\ie}{{\em i.e.\ }}

\newcommand{\eqcomma}{\laspace,}
\newcommand{\eqstop}{\laspace.}

\newcommand{\vc}[1]     {\ensuremath{\boldsymbol{#1}}}
\newcommand{\true}[1]   {\ensuremath{\tilde{#1}}}

\newcommand{\p}         {\ensuremath{p}}
\newcommand{\vp}        {\vc{\p}}

\newcommand{\q}         {\ensuremath{q}}
\newcommand{\vq}        {\vc{\q}}
\newcommand{\po}        {{\p_0}}
\newcommand{\vx}        {\vc{x}}

\newcommand{\y}         {\ensuremath{y}}

\newcommand{\usigma}        {\vc{\sigma}}

\newcommand{\truep}     {\true{\p}}
\newcommand{\vtruep}    {\true{\vp}}
\newcommand{\truedp}    {\ensuremath{\Delta\true{\p}}}
\newcommand{\trueq}     {\true{\q}}
\newcommand{\vtrueq}    {\true{\vq}}
\newcommand{\truepo}    {\true{\p}_0}

\newcommand{\trueind}     {\vc{\true{\y}}}

\renewcommand{\time}    {\ensuremath{t}}
\newcommand{\vtime}     {\vc{\time}}
\newcommand{\logt}      {\ensuremath{\tau}}
\newcommand{\fptime}    {\ensuremath{T}}
\newcommand{\vfptime}   {\vc{\fptime}}

\newcommand{\np}        {\ensuremath{m}}
\newcommand{\nq}        {\ensuremath{N}}
\newcommand{\nregion}        {\ensuremath{n}}

\newcommand{\g}         {\ensuremath{g}}
\newcommand{\z}         {\ensuremath{z}}

\newcommand{\pars}      {\vc{\phi}}

\newcommand{\var}       {\ensuremath{\sigma^2}}
\newcommand{\varp}      {\ensuremath{\var_\p}}
\newcommand{\varq}      {\ensuremath{\var_\q}}
\newcommand{\varpo}     {\ensuremath{\var_\po}}

\newcommand{\precis}       {\ensuremath{\sigma^{-2}}}
\newcommand{\precisp}      {\ensuremath{\precis_\p}}
\newcommand{\precisq}      {\ensuremath{\precis_\q}}
\newcommand{\precispo}     {\ensuremath{\precis_\po}}

\newcommand{\sd}       {\ensuremath{\sigma}}
\newcommand{\sdp}      {\ensuremath{\sd_\p}}
\newcommand{\sdq}      {\ensuremath{\sd_\q}}
\newcommand{\sdpo}     {\ensuremath{\sd_\po}}

\newcommand{\cfn}       {\vc{c}}
\newcommand{\cmat}      {\vc{C}}

\newcommand{\middlemid}{\isp\vert\isp} 
\newcommand{\given}{\middlemid}

\newcommand{\know}{\kappa}

\newcommand{\N}{\text{N}}
\newcommand{\Ga}{\text{Ga}}
\newcommand{\U}{\text{U}}

\makeatletter
\renewcommand\fps@figure{htbp}
\makeatletter

\newcommand{\figref}[1]{Figure~\ref{#1}}
\newcommand{\secref}[1]{Section~\ref{#1}}
\newcommand{\tabref}[1]{Table~\ref{#1}}

\title[Preprint]{A Bayesian approach to deconvolution in well test analysis}
\author[T. Botsas, I. H. Jermyn and J. A. Cumming]{Themistoklis Botsas, Jonathan A. Cumming and Ian H. Jermyn}
\address{Department of Mathematical Sciences,
        Durham University,
        Durham,
        UK.}
\email{tbotsas@turing.ac.uk}

\begin{document}
\begin{abstract}


In petroleum well test analysis, deconvolution is used to obtain information about the reservoir system. This information is contained in the response function, which can be estimated by solving an inverse problem in the pressure and flow rate measurements. Our Bayesian approach to this problem is based upon a parametric physical model of reservoir behaviour, derived from the solution for fluid flow in a general class of reservoirs. This permits joint parametric Bayesian inference for both the reservoir parameters and the true pressure and rate values, which is essential due to the typical levels of observation error. Using a set of flexible priors for the reservoir parameters to restrict the solution space to physical behaviours, samples from the posterior are generated using MCMC. Summaries and visualisations of the reservoir parameters' posterior, response, and true pressure and rate values can be produced, interpreted, and model selection can be performed. The method is validated through a synthetic application, and applied to a field data set. The results are comparable to the state of the art solution, but through our method we gain access to system parameters, we can incorporate prior knowledge that excludes non-physical results, and we can quantify parameter uncertainty.

\end{abstract}


\section{Introduction}

Within petroleum engineering, well test analysis encompasses a set of methodologies for the planning, conducting, and interpretation of the results of controlled experiments on a well in a petroleum reservoir. The well test experiment controls the flow rate of fluid from the reservoir and observes the pressure at the bottom of the well (the `bottomhole pressure'), producing a pair of time series. The goal of well test analysis is then to identify and interpret the relationship between pressure and rate, and use this to make inferences about reservoir parameters, properties, geology, and geometry \citep{bourdet2002well}.

A variety of technical tools have been developed in order to analyse the data from well tests \citep{gringarten2008straight}, such as straight line analyses \citep{horner1951pressure}, type curve matching \citep{gringarten1979comparison}, and derivative curve analysis \citep{bourdet1983new}. However, these standard methods are typically restricted to the analysis of periods of constant-rate data, due to the corresponding simplification of the pressure-rate relationship under such conditions. Hence, variable rate data has to be treated as a collection of smaller independent data sets, resulting in inconsistencies in the analysis arising from the obvious dependency induced by the cumulative effect of production over time.

Deconvolution emerged as a major milestone in well test analysis \citep{von2004deconvolution} due to its ability to deal with variable rate tests in their entirety, enabling the consistent analysis of larger and richer data than was previously accessible. The ability to deconvolve entire pressure and rate histories enables a coherent analysis of longer-term behaviour, providing greater insights into the flow behaviour of the reservoir and more distant geological features. 

The key relationship for this inference is that, under appropriate regularity conditions and ranges of reservoir conditions, the bottomhole pressure, $\truep$, and the flow rate, $\trueq$, written as functions of time $\time$, are accurately related by:
\begin{equation}
 \truedp(\time)  = 
        \truepo - \truep(\time) 
        = \int_{0}^{\infty} \isp \g(\time - \time') \isp \trueq(\time') \isp d\time' 
        = \int_{0}^{\infty} \isp \trueq(\time - \time') \isp \g(\time') \isp d\time' ,  
        \label{eq:Deconv}
 \end{equation}
 
\noi where $\truedp$ is the pressure drop from an initial equilibrium pressure $\truepo$ at $\time = 0$, and use has been made of the fact that $\g(t) = \trueq(t) = 0$ for $\time < 0$, by causality and by definition respectively. The \emph{reservoir response function} $\g$ is the object of primary interest, because it encapsulates all relevant well- and reservoir-specific information in a single function, and thus provides an effective summary and signature for the behaviour of a particular well in response to production. The value $\g(\Delta\time)$ describes the effect on the pressure at $\time$ of the rate at $\time - \Delta\time$; it is also the derivative of the pressure drop at time $\Delta\time$ induced by a unit flow from $\time = 0$. In principle, it can be derived as the coincidence limit of the Green function of the diffusion equation with $\trueq$ as a time-varying point source and with diffusivity and mobility coefficients spatially-varying functions of reservoir and well properties.

\begin{figure}
\begin{center}
\begin{tikzpicture}[scale=0.34]
    \node[inner sep=0pt] (pres) at (-30,0)
   {\includegraphics[width=.3\textwidth]{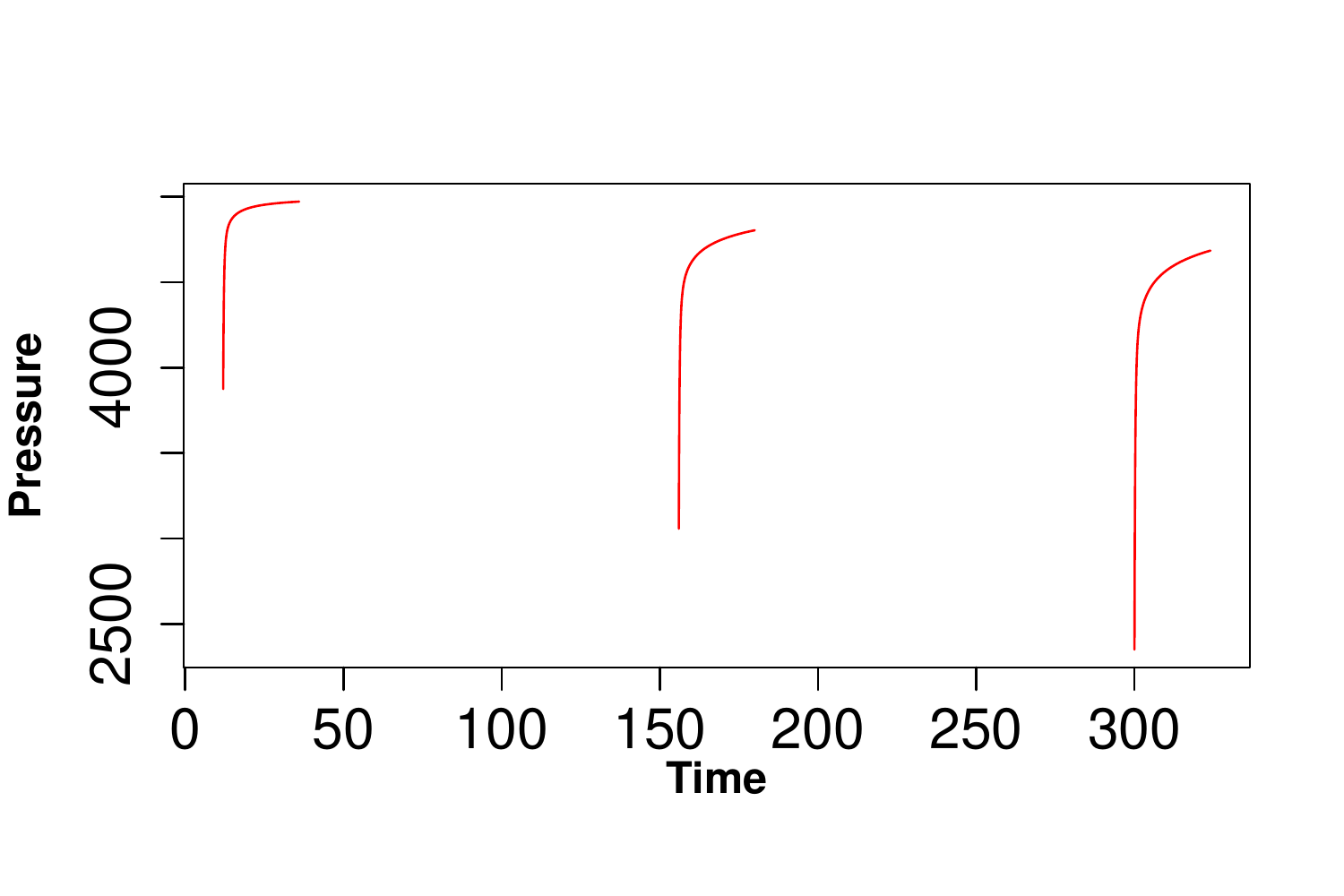}};

   	\draw [->,thick] (-9,0) -- (-7,0) node[midway,above]{};
    \node[inner sep=0pt] (rate) at (-15,0)
   {\includegraphics[width=.3\textwidth]{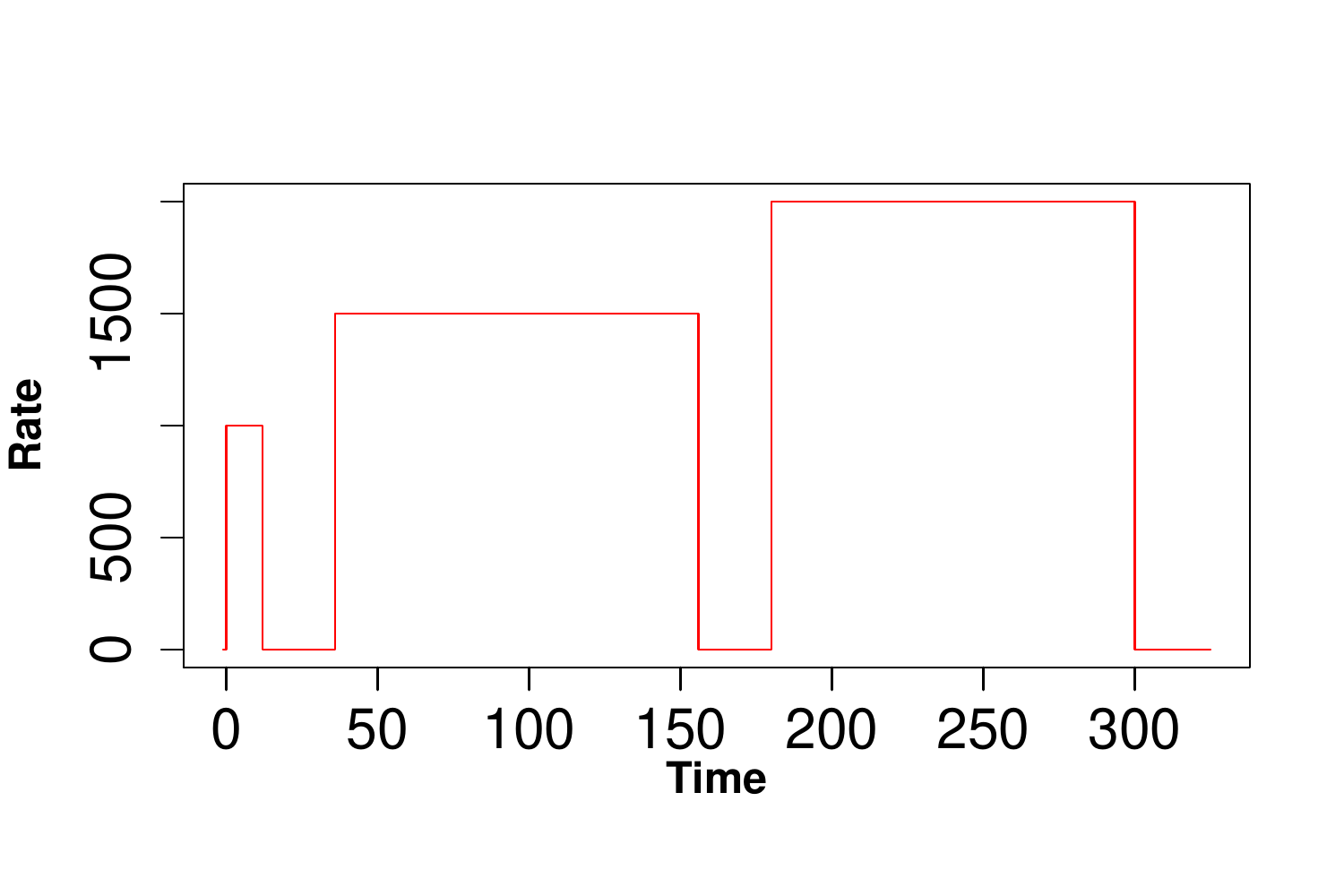}};
	\node[inner sep=0pt] (resp) at (0,0)
    {\includegraphics[width=.3\textwidth]{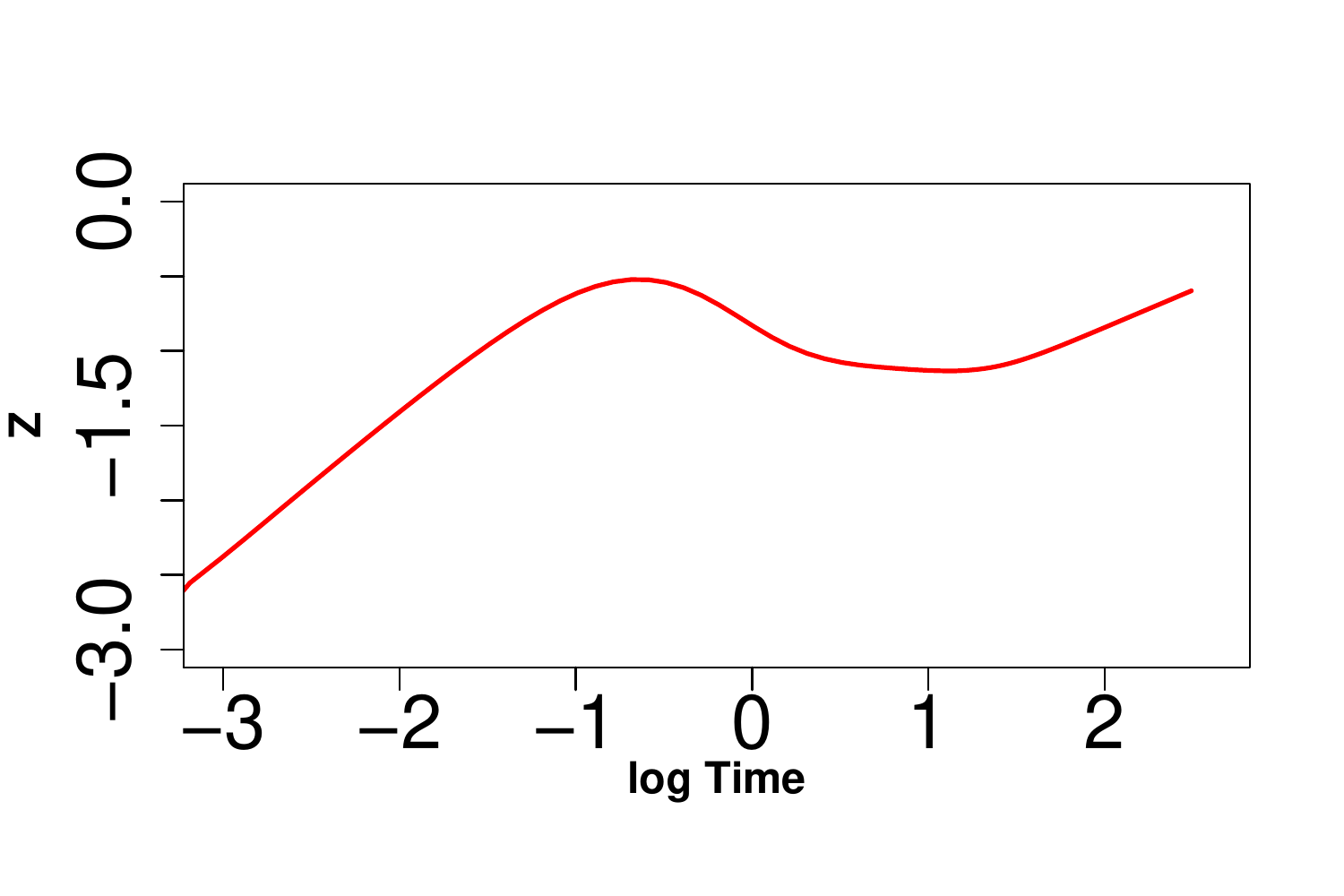}};
    \node[] (plus) at (-22.5,0) {\textbf{+}};
\end{tikzpicture}

    \caption{The deconvolution of the bottomhole pressure, $\truep$ (left), and the flow rate, $\trueq$ (middle) produce the reservoir response function $\g$, whose transformation $\z$ (right) is plotted here.}\label{deconv_fig}
\end{center}
\end{figure}

For visualisation, it is better to plot $\z(\logt)=\logt + \ln\g(e^\logt)$, where $\logt=\ln \time$ \citep{bourdet1983new}; \figref{deconv_fig} shows an example. This greatly facilitates human interpretation of the response function, with features of the plot being directly associated with particular flow regimes and reservoir features. These features may be categorised by the times at which they appear, where time is effectively a surrogate for distance from the wellbore.
\begin{itemize}

\item 

Early-time (a few seconds or minutes) is dominated by the wellbore and its immediate surroundings. The response usually begins with a unit slope straight line, representing the effect of fluid filling the uniform wellbore, typically followed by a characteristic `bump', which represents the impedance caused by localised damage due to drilling, known as the `skin'.

\item

Middle-time (a few hours or days) usually shows stabilization as the well draws fluid from the relatively homogeneous region beyond the damaged skin, where a `radial flow' regime is established in which fluid moves towards the well from all directions.

\item

Late-time (weeks or months) indicates features of the reservoir far from the wellbore, such as impermeable boundaries or other heterogeneities. 

\end{itemize}

\begin{figure}
\centering
\begin{subfigure}[]{0.49\columnwidth}
	\begin{tikzpicture}[scale=0.3]
		\draw [->,thick] (1,0) -- (8,0) node[midway,above]{};
		\fill [yellow] (-2,-2) rectangle (2,2);
		\draw[fill] (0,0) circle [radius=2pt];
		\node[inner sep=0pt] (mixed) at (12,0)
		{\includegraphics[width=.4\textwidth]{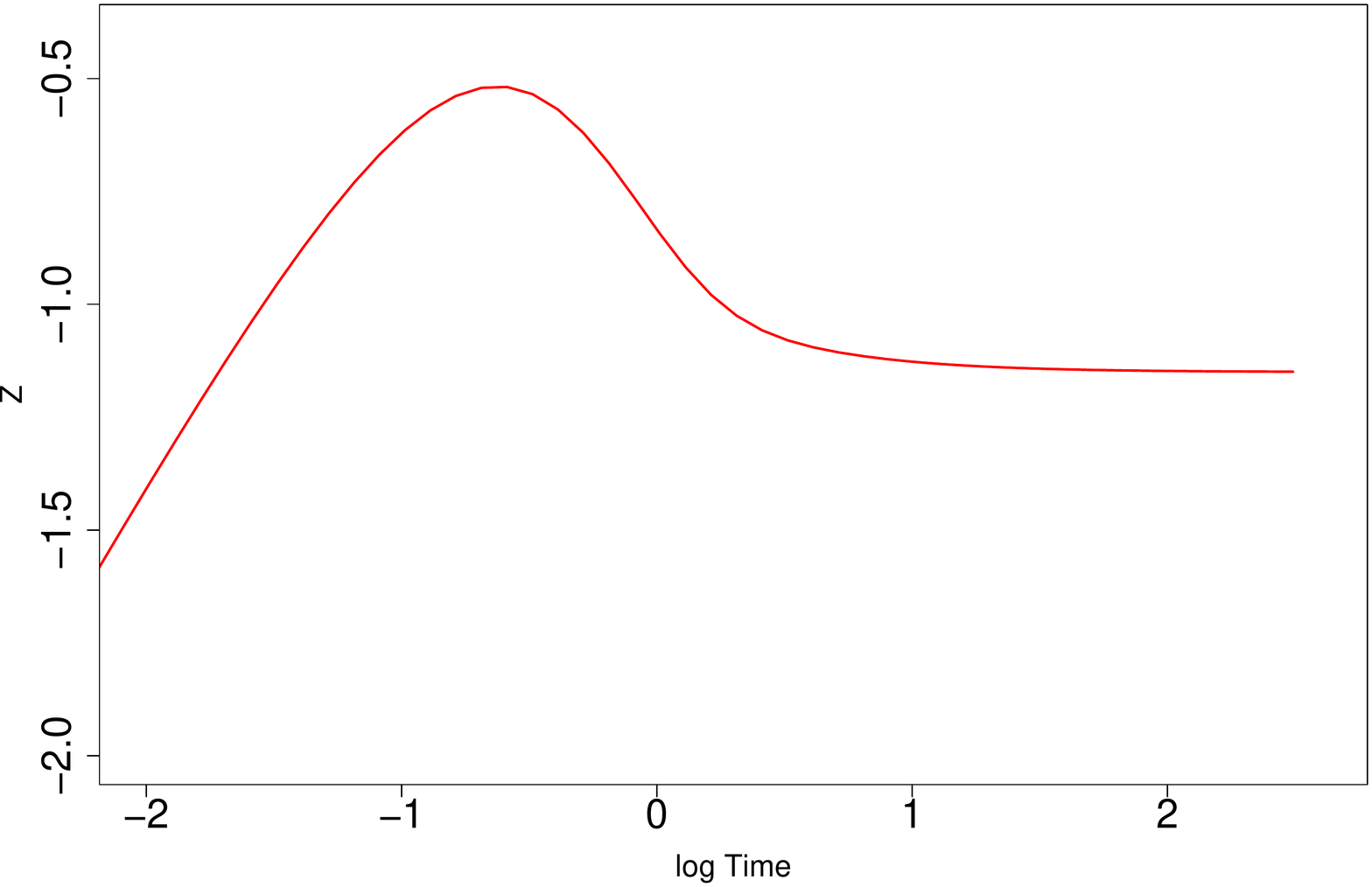}};
	\end{tikzpicture}
    \caption{Infinite radial flow (unbounded reservoir).}

	\begin{tikzpicture}[scale=0.3]
		\draw[thick] (-2,-2) -- (2,-2);
		\draw [->,thick] (1,0) -- (8,0) node[midway,above]{};
		\fill [yellow] (-2,-2) rectangle (2,2);
		\draw[fill] (0,0) circle [radius=2pt];
		\node[inner sep=0pt] (mixed) at (12,0)
		{\includegraphics[width=.4\textwidth]{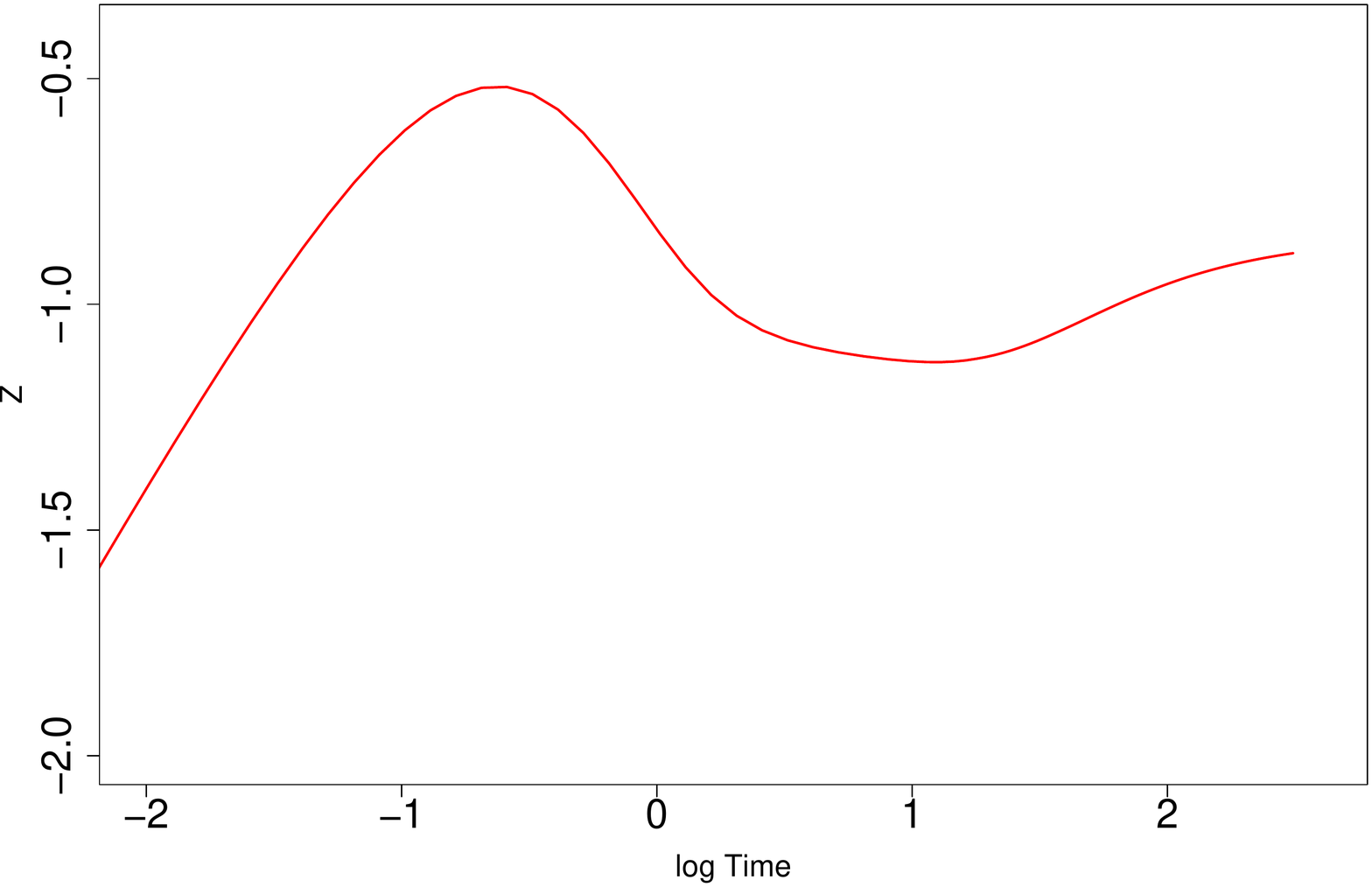}};
	\end{tikzpicture}
    \caption{Single fault.}
\end{subfigure}
\begin{subfigure}[]{0.49\columnwidth}
	\begin{tikzpicture}[scale=0.35]
		\draw [->,thick] (1,0) -- (8,0) node[midway,above]{};
		\draw[thick] (-2,-2) -- (2,-2);
		\draw[thick] (-2,2) -- (2,2);
		\fill [yellow] (-2,-2) rectangle (2,2);
		\draw[fill] (0,0) circle [radius=2pt];
		\node[inner sep=0pt] (mixed) at (12,0)
		{\includegraphics[width=.4\textwidth]{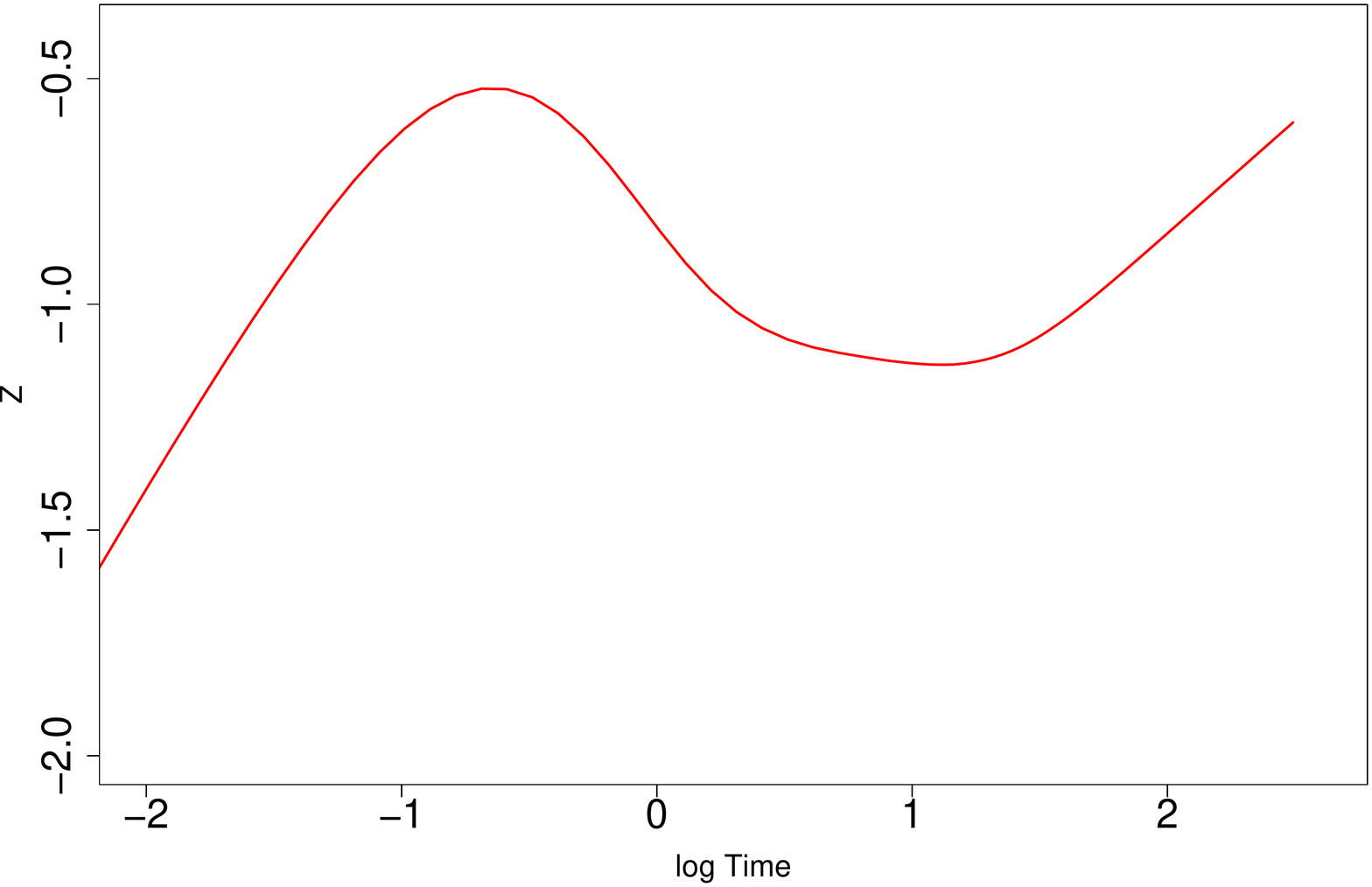}};
	\end{tikzpicture}
    \caption{Channel (two parallel faults).}

	\begin{tikzpicture}[scale=0.35]
		\draw [->,thick] (1,0) -- (8,0) node[midway,above]{};
		\draw[thick] (-2,-2) -- (2,-2);
		\draw[thick] (-2,2) -- (2,2);
		\draw[thick] (-2,-2) -- (-2,2);
		\draw[thick] (2,-2) -- (2,2);
		\fill [yellow] (-2,-2) rectangle (2,2);
		\draw[fill] (0,0) circle [radius=2pt];
		\node[inner sep=0pt] (mixed) at (12,0)
		{\includegraphics[width=.4\textwidth]{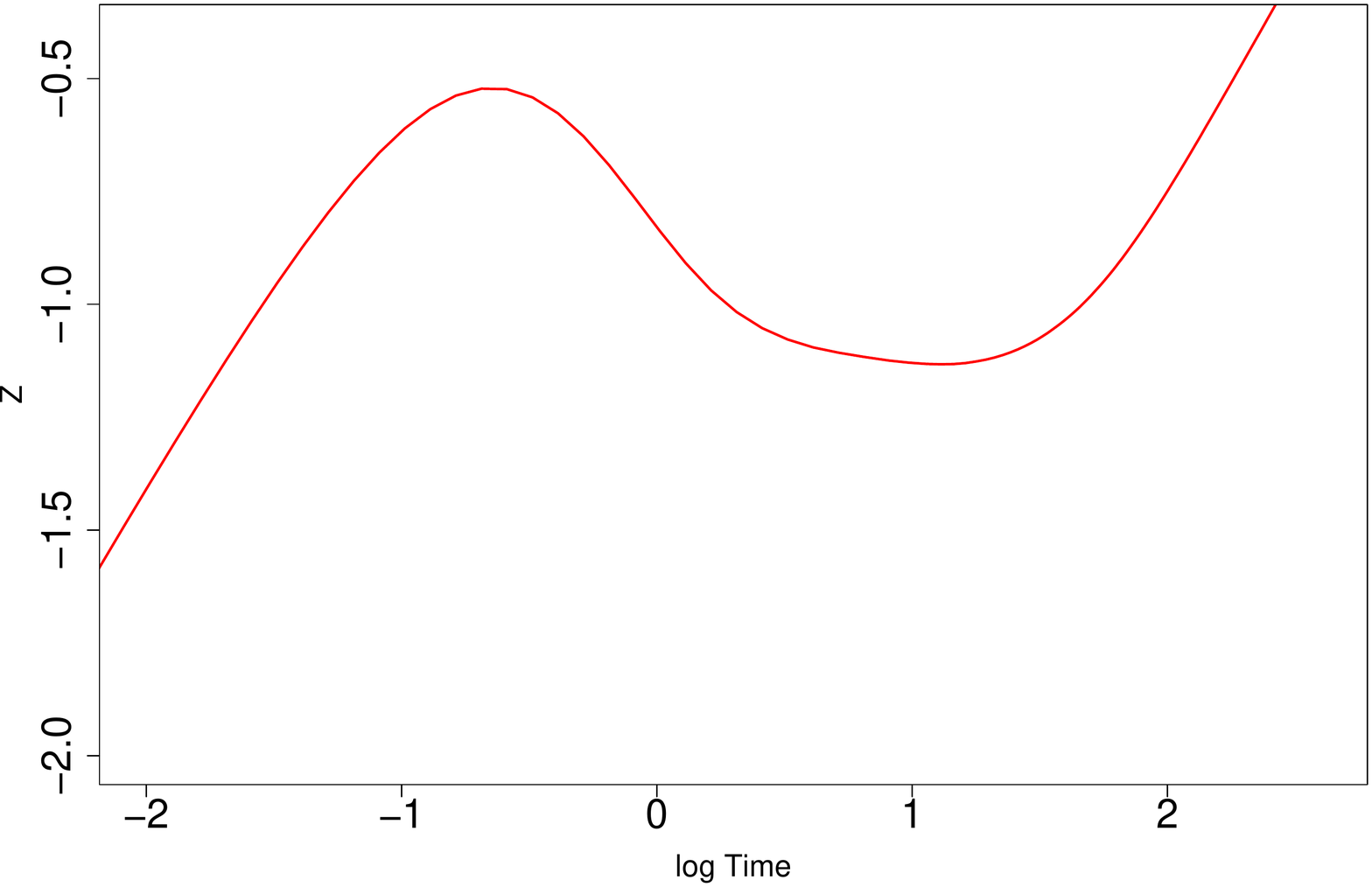}};
	\end{tikzpicture}
    \caption{Closed reservoir (depletion).}
    \end{subfigure}
    \caption{Top view of indicative reservoir configurations (left subfigures) and resulting response functions (right subfigures).}\label{reserv_shapes}
    \end{figure}

\figref{reserv_shapes} shows some common response shapes. Note that a specific shape may not uniquely identify a reservoir feature (\eg the fault response can also be achieved by appropriate changes in mobility) and that sequential features may mask each other.

Of course, we do not observe entire pressure and rate functions: we only observe values at a finite set of times. Furthermore, these observations do not give values of $\truep$ or $\trueq$ directly, but rather values subject to measurement error and other uncertainties. \citet{von2004deconvolution} introduced the first deconvolution method that took these issues into account using a form of penalized total least squares (TLS) regression, with the response $\z$ represented as a piecewise linear function. The method has a number of limitations. First, it requires, and is highly sensitive to, the specification of multiple hyperparameters; second, regularization is required to produce a smooth response function; third, the flexibility of the linear spline representation means that the resulting response functions are not guaranteed to be physically possible; and fourth, it lacks a coherent approach to uncertainty analysis.

\subsection{Problem formulation}

To address these limitations, we propose a Bayesian approach. First, we replace the linear spline representation of the response function by a parametric model based on an explicit yet sufficiently general solution to the diffusion equation, thus avoiding the need for regularization while restricting attention to physically sensible responses. Second, the aforementioned hyperparameters are treated as nuisance parameters and integrated out. Finally, the Bayesian approach inherently provides for a coherent uncertainty analysis.

We first establish some notation. The well pressure, $\truep$, is observed at times $\vtime = \{\time_{i}\}$, idealised as point measurements, giving pressure data $\vp = \{\p_{i}\}$, for $i= 1,\dots,\np$. We have a single observation, $\po$, for the initial reservoir pressure $\truepo$. The rate is modelled as periods of constant flow, $\vtrueq = \{\trueq_{j}\}$, defined over known time intervals $\vfptime = \{[\fptime_j,\fptime_{j+1}]\}$, which are observed as $\vq = \{\q_{j}\}$, for $j=1,\dots,\nq$. We denote the piecewise constant function given by $\vfptime$ and $\vtrueq$, with zeroes elsewhere, as $\trueq$. 

We are interested in the probability $P(\g, \vtrueq, \truepo \given \vp, \vq, \po, \know)$, where $\know$ represents any other prior knowledge \eg $\vfptime$. Bearing in mind equation~\eqref{eq:Deconv}, we have
    \begin{multline}
        P(\g, \vtrueq, \truepo \given \vp, \vq, \po, \know)
        \propto
        P(\vp \given \truepo, \g*\trueq, \know)
        \isp P(\vq \given \vtrueq, \know) \\
        \times P(\po \given \truepo, \know)
        \isp P(\g \given \know)\isp P(\vtrueq \given \know)\isp P(\truepo \given \know)
        \eqcomma
        \label{eq:probabilitybreakdown}
    \end{multline}

\noi where $*$ indicates convolution, and we have used various independences to be described later. As stated, $\g$ will be given by a parametric model based on an explicit solution to the diffusion equation, with parameters $\pars$, so that the distributions on $\g$ will in fact be on $\pars$. In the next section, we describe this model in detail.

\section{The response model} \label{sec:respmodel}

The main challenge for a Bayesian approach to well test analysis is to identify an appropriate prior distribution for the response function. The model should be flexible enough to capture the shapes encountered in practice, restrictive enough to prohibit non-physical results, and computationally tractable. In addition, a model expressed in terms of well and reservoir properties would support more detailed interpretation than is possible with a generic representation such as a spline, permitting comparison to the results of other well test analyses; providing constraints on, and a physical interpretation for, the prior; and perhaps providing direct information about physical aspects of the system.

Here we use a variation of the `multi-region radial composite reservoir model'~\citep{acosta1994thermal, zhang2010well} to satisfy these desiderata. First, through appropriate choice of parameter values, the response function can be made to resemble the majority of plausible response shapes. Second, since the model is derived as a solution to the diffusion equation representing the physical fluid flow problem, it is restricted to physically sensible forms. Third, the finite-dimensional nature of the model makes it far more computationally tractable than a description of an arbitrary reservoir. Finally, the model itself is parameterized in a way that can be associated with the flow behaviour in the reservoir~\citep{bourdet2002well}.
 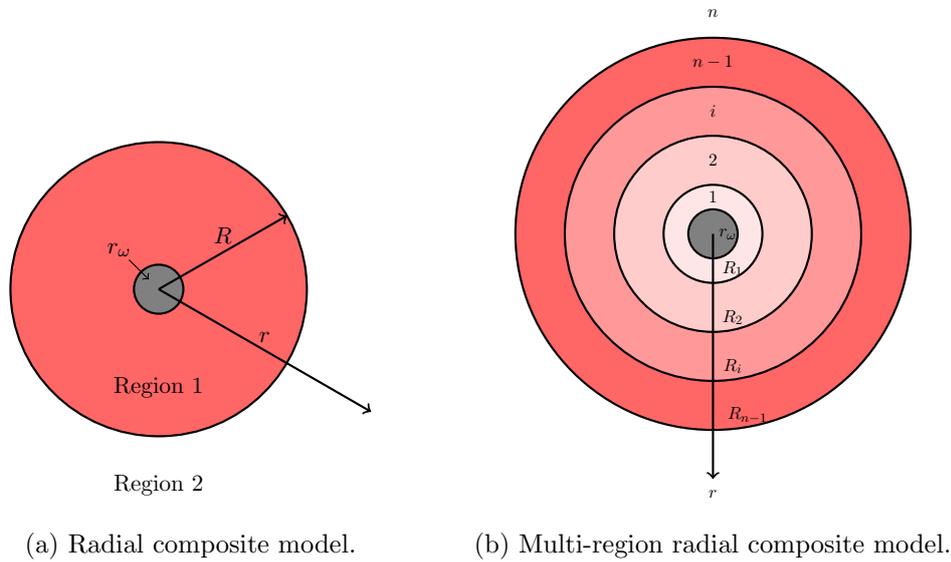
\begin{figure}
 \centering
 \begin{subfigure}[t]{0.45\textwidth}
 \centering
	\begin{tikzpicture}[scale=0.65,
	every node/.style={scale=0.8}]
		\draw [thick, fill = red!60] circle [radius=3];
		\draw [fill=gray, thick] circle [radius=0.5];
		\draw [->,thick] (0,0) -- (2.6,1.5) node[midway,above]{$R$};
		\draw [->,thick] (0,0) -- (4.3,-2.5) node[midway,above]{$r$};
		\node[] at (0,-2) {Region 1};
		\node[] at (0,-4) {Region 2};
		\node[] (rw) at (-0.8,0.8) {$r_\omega$};
		\draw [->] (-0.6,0.6) -- (-0.2,0.2);
	\end{tikzpicture}
	\caption{Radial composite model.} \label{rcm}
\end{subfigure}%
~
\begin{subfigure}[t]{0.45\textwidth}
\centering
 	\begin{tikzpicture}[scale=0.65,
 	    every node/.style={scale=0.8}]
		\draw [thick, fill = red!60] circle [radius=4];
		\draw [thick, fill = red!40] circle [radius=3];
		\draw [thick, fill = red!20] circle [radius=2];
		\draw [thick, fill = red!10] circle [radius=1];
		\draw [fill=gray, thick] circle [radius=0.5];
		\node[scale=0.75] at (0,4.5) {$n$};
		\node[scale=0.75] at (0,3.5) {$n-1$};
		\node[scale=0.75] at (0,2.5) {$i$};
		\node[scale=0.75] at (0,1.5) {$2$};
		\node[scale=0.75] at (0,0.75) {$1$};

		\node[scale=0.75] at (0.3,0) {$r_\omega$};
		\draw [->,thick] (0,0) -- (0,-5);
		\node[scale=0.75] at (0.4,-0.7) {$R_1$};
		\node[scale=0.75] at (0.4,-1.7) {$R_2$};
		\node[scale=0.75] at (0.4,-2.7) {$R_i$};
		\node[scale=0.75] at (0.7,-3.7) {$R_{n-1}$};
		\node[scale=0.75] at (0,-5.3) {$r$};
	\end{tikzpicture}
	\caption{Multi-region radial composite model.} \label{mrrcm}
	\end{subfigure}
	\caption{The radial composite reservoir models represent the reservoir as two (left) or more (right) concentric circular regions, each with different, constant mobility and diffusivity values, in the centre of which lies the wellbore.}
 \end{figure}

The multi-region radial composite model is based on the simple radial composite model with an infinite outer region~\citep{satman1980interpretation}. This model represents the reservoir as two concentric circular regions (the second region tends to infinity), at the centre of which lies the wellbore, as shown in \figref{rcm}. The model has six parameters: three correspond to a well in a homogeneous reservoir~\citep{bourdet1983new}, and affect early-time behaviour and global scales; the remaining three characterize the transition at the boundary of the inner region, and primarily affect the mid- to late-time behaviour. These parameters are sufficient to define the diffusion equation for the system, and hence to derive the response function. They are summarised in \tabref{tab:params}. 

\begin{table}
    \centering
    \caption{Parameters of the radial composite reservoir model.}
    \begin{tabular}{c|p{2.75cm}|l|p{3cm}|l}\hline
         Parameter & Name & Scope & Description & Definition  \\ 
         \hline
         $P$ & Pressure match & Global & Pressure scale & 
         $P = \log\left(2 \pi h (k/\mu)_1\right)$
         \\
         $T$ & Time match & Global & Time scale & 
         $T = \log\left({\frac{2 \pi h (k/\mu)_1}{C}}\right)$
         \\ 
         \hline
         $W$ & Wellbore storage coefficient & Early time & Early time shape & 
         $W = \log\left({\frac{C e^{2 S}}{2 \pi \phi c_t h r_w^2}}\right)$
         \\ 
         \hline
         $R$ & Radius parameter & Transition & Dimensionless distance to transition & 
         $R = \log\left({\frac{r_1}{r_w e^{-S}}}\right)$
         \\
          $M$ & Mobility ratio & Transition & Ratio of mobility between regions & 
          $M=\log\left({\frac{(k/\mu)_1}{(k/\mu)_2}}\right)$
          \\
          $\eta$ & Diffusivity ratio & Transition & Ratio of diffusivity between regions & $\eta=\log\left({\frac{(k/\mu \phi c_t)_1}{(k/\mu \phi c_t)_2}}\right)$
          \\
         \hline
    \end{tabular}
    \label{tab:params}
\end{table}

The parameters can be written as functions of the physical properties of the wellbore and reservoir: the wellbore radius $r_w$, the wellbore storage coefficient $C$, the wellbore skin $S$, the formation thickness $h$, the total system compressibility $c_t$, the viscosity of the fluid $\mu$, the permeability $k$ and porosity $\phi$ of the medium, and the transition radius $r_1$. The relationships between the derived model parameters and these fundamental parameters are given in the final column of \tabref{tab:params}. The parameters $k$, $\mu$, $\phi$, and $c_t$ are piecewise constant over the regions of the radial model. Changes in their values at the transition boundaries represent reservoir features such as faults and changes in rock properties; due to the resulting changes in the six derived parameters, these then manifest themselves as identifiable features in the response function. However, the fundamental parameters are non-identifiable, and in the absence of strong prior information about their values, we will work here with the six derived parameters $\pars = (P, T, W, R, M, \eta )$. 

\begin{figure}
\centering
\begin{subfigure}{.33\textwidth}
  \includegraphics[width=\textwidth]{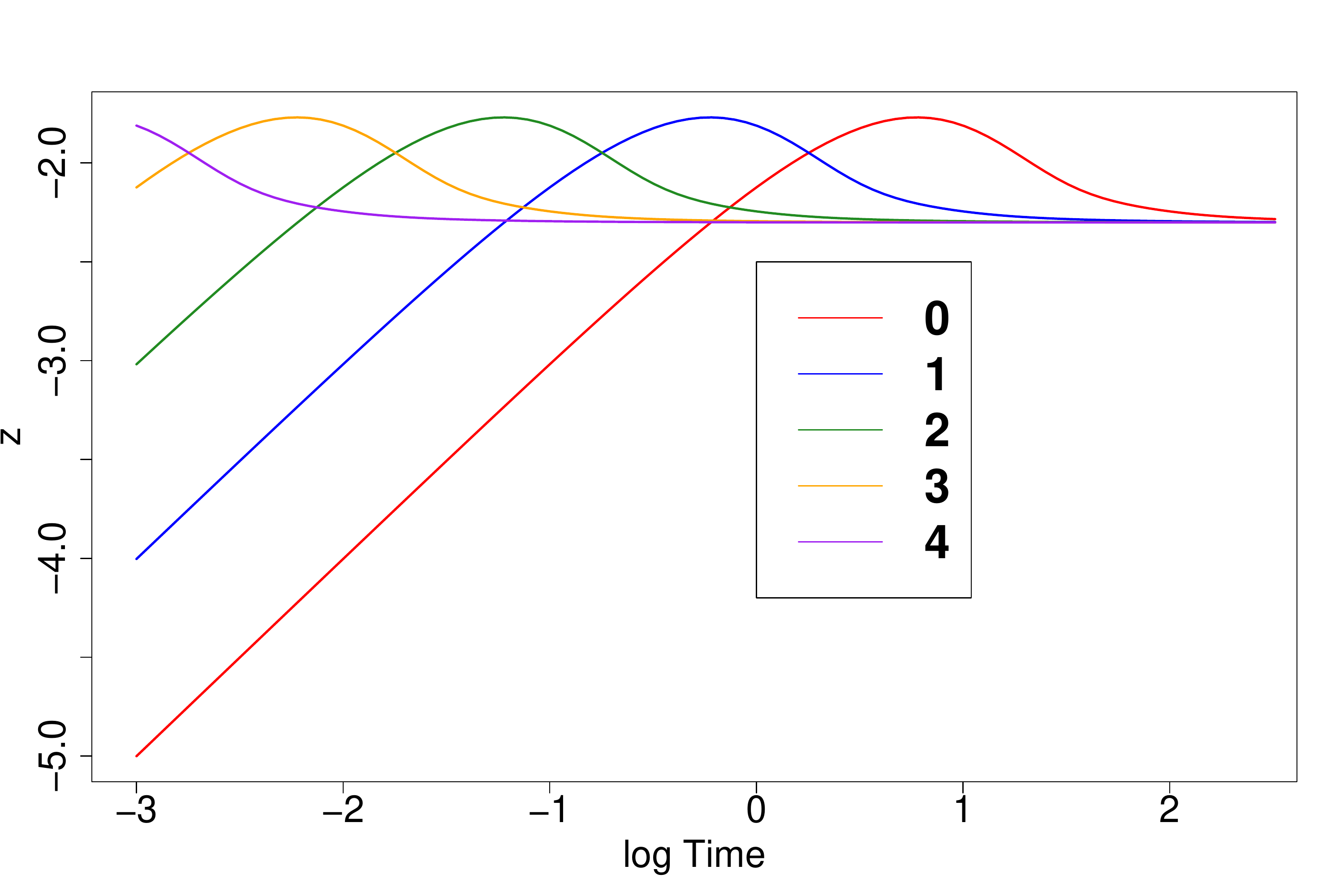}
  \caption{$T$}
  \label{fig:TM}
\end{subfigure}%
\begin{subfigure}{.33\textwidth}
  \includegraphics[width=\textwidth]{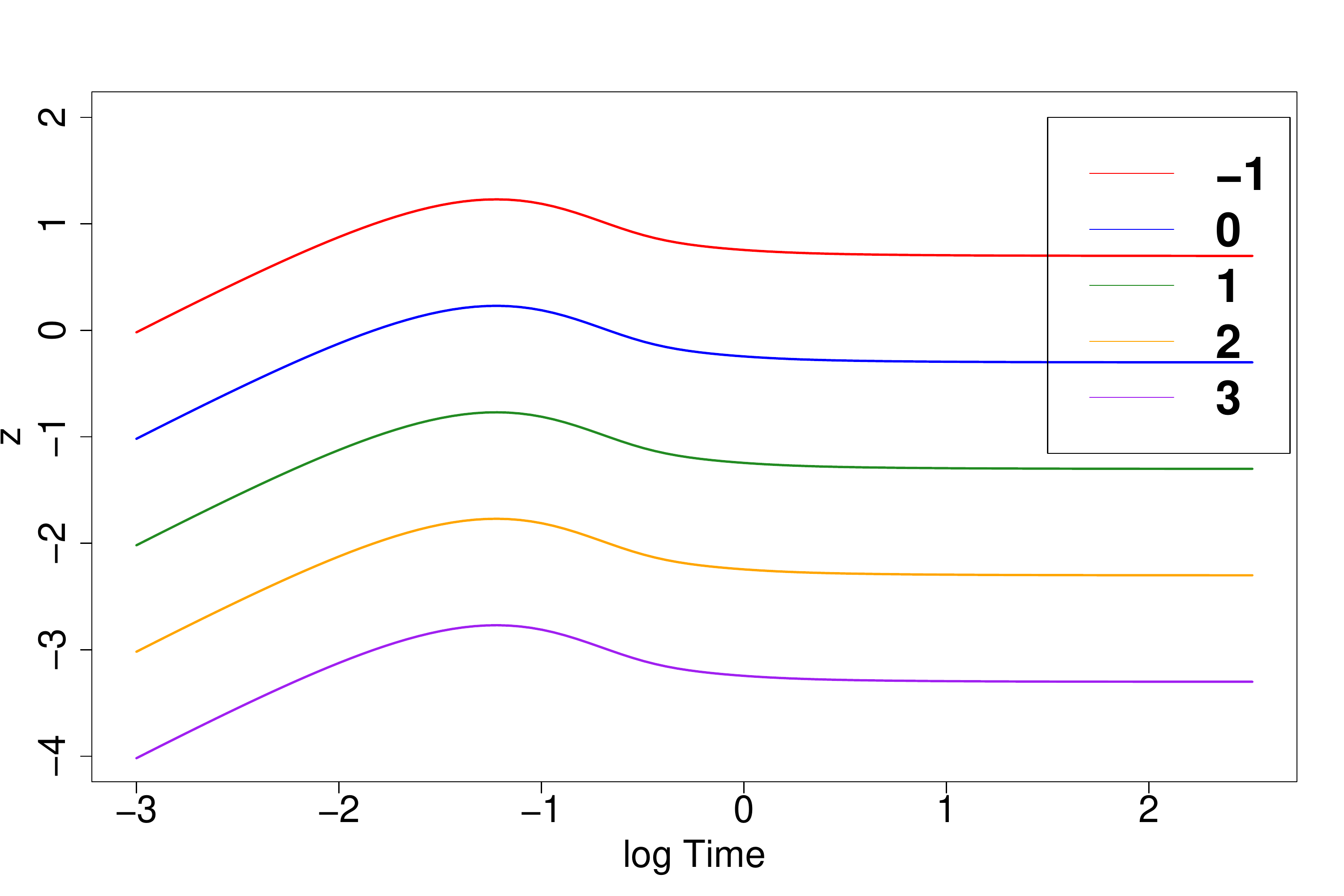}
  \caption{$P$}
  \label{fig:PM}
\end{subfigure}%
\begin{subfigure}{.33\textwidth}
  \includegraphics[width=\textwidth]{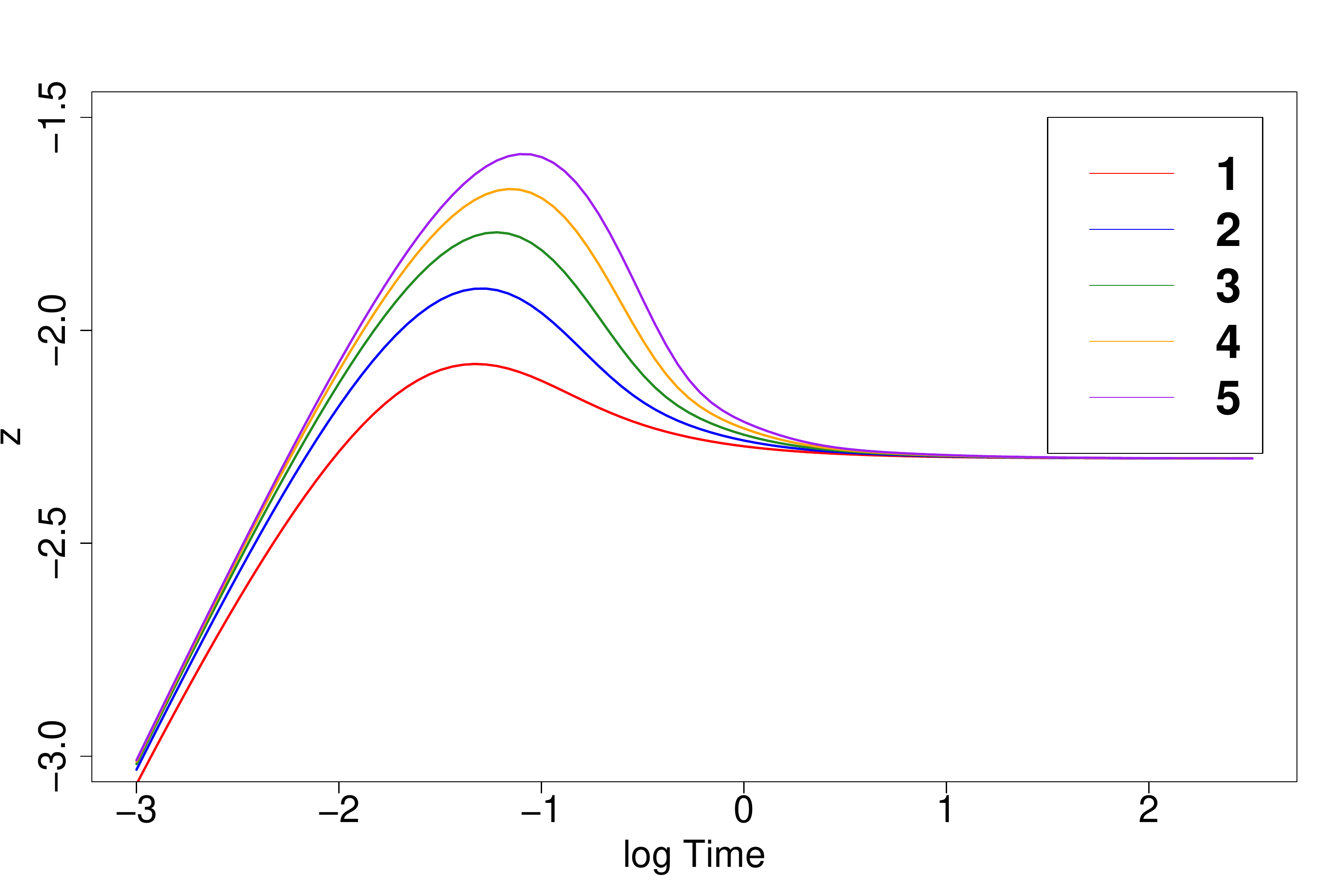}
  \caption{$W$}
  \label{fig:CDe2Skin}
\end{subfigure}\\
\begin{subfigure}{.33\textwidth}
  \includegraphics[width=\textwidth]{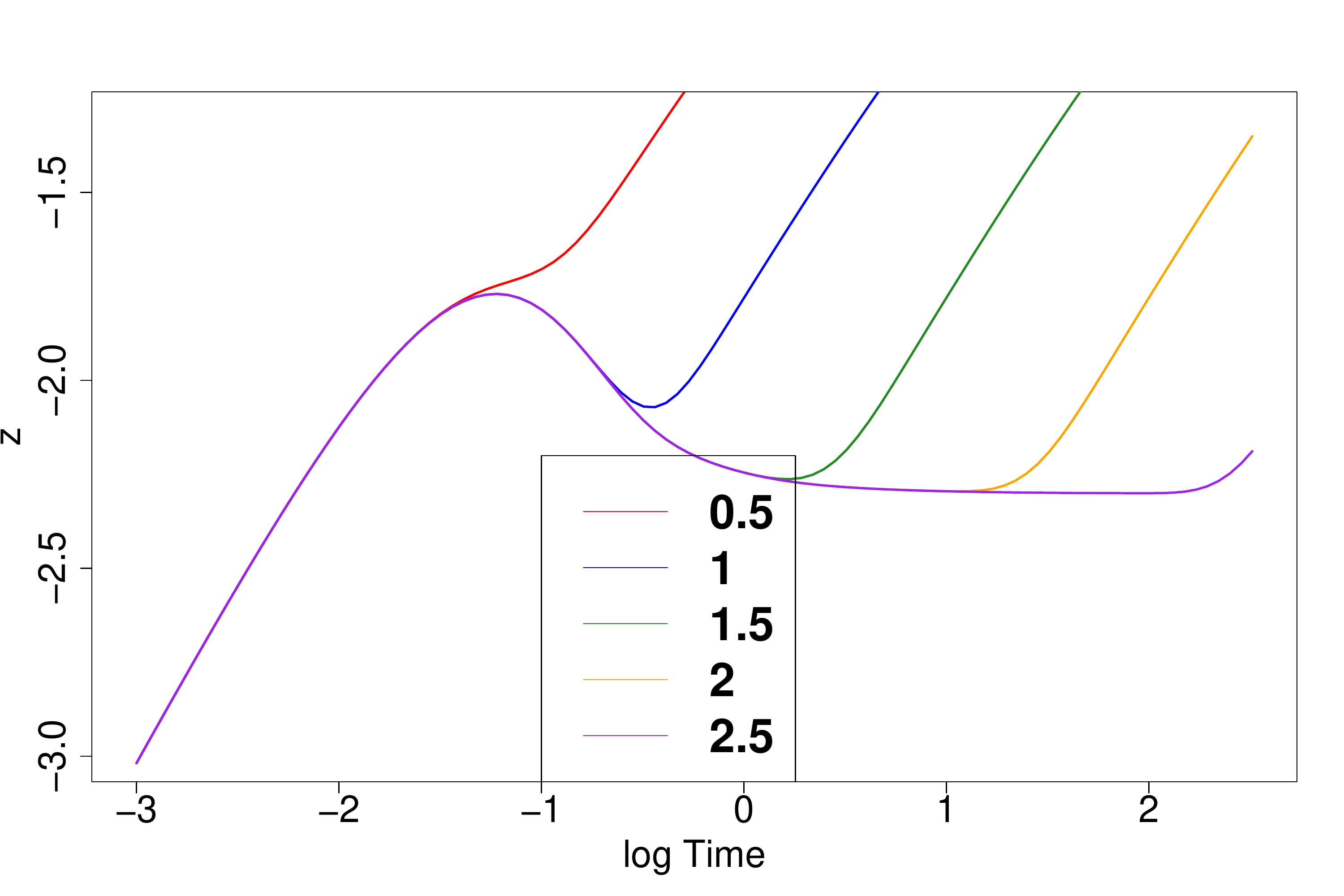}
  \caption{$R$}
  \label{fig:RD}
\end{subfigure}%
\begin{subfigure}{.33\textwidth}
  \includegraphics[width=\textwidth]{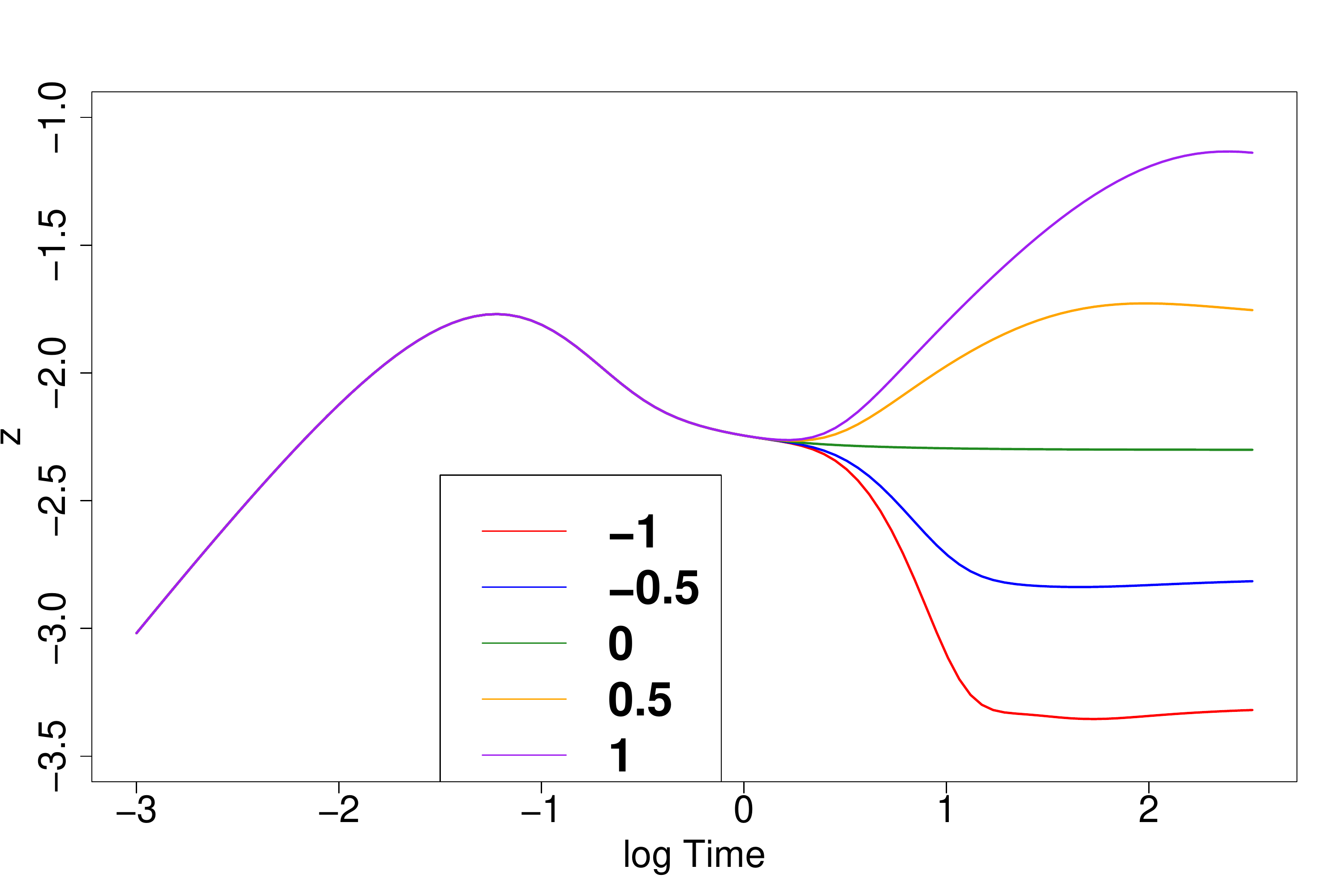}
  \caption{$M$}
  \label{fig:M}
\end{subfigure}%
\begin{subfigure}{.33\textwidth}
  \includegraphics[width=\textwidth]{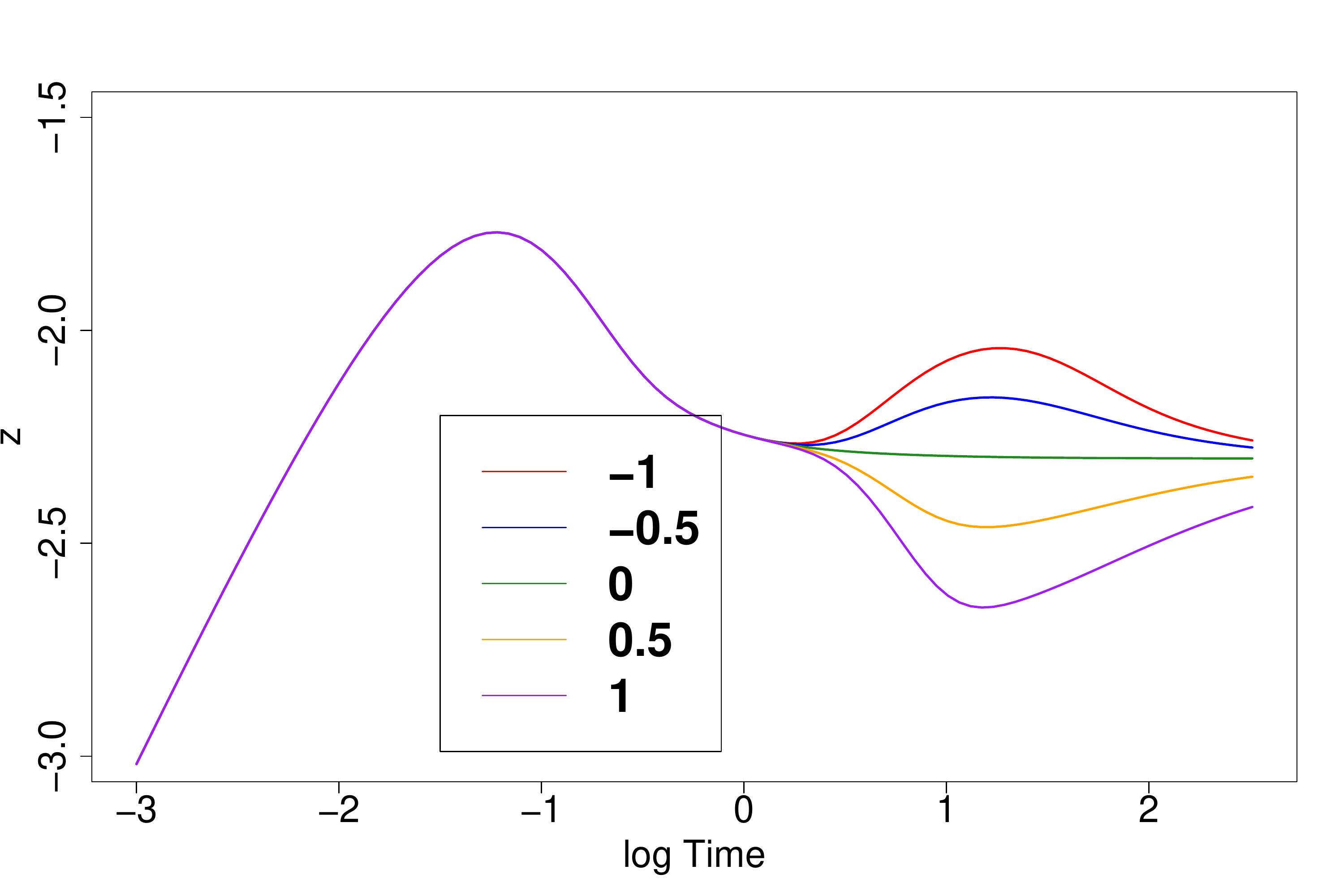}
  \caption{$\eta$}
  \label{fig:eta}
\end{subfigure}
\caption{Response function sensitivity to changes in the parameters $\pars$.}
\label{fig:rcm_params}
\end{figure}

\figref{fig:rcm_params} shows the effects on the features of the response function of changes in the parameters $\pars$. The parameters $T$ and $P$ define time and pressure scales, and represent the responsiveness of the well and reservoir to production. The parameter $W$ governs the impedance to flow due to the skin effect surrounding the well. Parameter $R$ corresponds to the distance from the wellbore to the inter-region transition, and so affects the time at which the impact of that transition is perceived. Finally, $M$ and $\eta$ describe the relative changes in reservoir properties at the transition, resulting in a shift in the response function stabilisation level, or a localised deviation from that level. In view of the multiplicative nature of the parameters $T$, $P$, $W$, $M$ and $\eta$, logarithms are taken (base 10 because this is the industry standard, in which values are most easily interpretable). The radius parameter $R$ determines the time at which features appear in the response function. In view of the similar spacing between response features when plotted on a log time scale, and of the relationship between time and distance from wellbore, the radius parameter is also treated on a log scale. 

The simple radial model can be extended to the multi-region case \citep{acosta1994thermal, zhang2010well} by introducing additional concentric radial regions around the model, resulting in a model of $\nregion$ regions and $\nregion-1$ transitions as shown in \figref{mrrcm}. 
The multi-region model adds to the parameters of the simple radial model an additional radius increment $R_{i} = \log\left({r_{i} - r_{i-1} \over r_{w}e^{-S}}\right)$, mobility ratio $M_{i} = \log\left({(k/\mu)_{i} \over (k/\mu)_{i+1}}\right)$, and diffusivity ratio $\eta_{i} = \log\left({(k/\mu\phi c_t)_{1} \over (k/\mu\phi c_t)_{i+1}}\right)$ for each transition beyond the first. This gives a total of $3n$ parameters: $\pars = (P, T, W, \{R_{i}, M_{i}, \eta_{i}\})$, with $i = 1,\dotsc, (n-1)]$. The extra regions and their parameters mean that the model can now describe complex combinations of the features shown in \figref{fig:rcm_params}: \eg a shift of the stabilisation level, followed by a localised deviation from that level, requires at least two transitions, where, in the former there is a change in mobility, and in the latter, in diffusivity.  However, it also raises the question of how many regions to use, which is addressed in \secref{sec:oil_data}. 

\section{The Bayesian model}\label{stat model}

Having described the model for the response function $\g$, we now describe the various components of equation~\eqref{eq:probabilitybreakdown}.

\subsection{The data model}\label{data_model}
We first consider $P(\vp \given \truepo, \g*\vtrueq, \know) = P(\vp \given \truepo, \pars, \vtrueq, \know)$, which represents the observational error in the pressure data. We introduce a variance parameter $\varp$, and use a multivariate Gaussian distribution for the observed pressures when $\varp$ is known in addition to the other parameters:
\begin{equation}\label{eq:pres_match}
    \vp \given \pars, \trueind, \varp, \know
    \sim N\left(
    \vtruep(\pars, \trueind), I_{\varp}
    \right)
    \eqstop
\end{equation}

\noi where $I_x$ denotes the diagonal matrix with $x$ on the main diagonal, $\trueind = (\vtrueq, \truepo)$, $\vtruep = \truepo - \cfn(\g(\pars)*\trueq)$, and $\cfn$ samples any continuous function of $\time$ at the points $\vtime$. The piecewise constant nature of $\trueq$ means that we can go further, and write $\cfn(\g(\pars)*\trueq) = \cmat(\pars) \vtrueq$, where $\cmat(\pars)$ is an $(\np\times\nq)$ matrix; this makes the convolution computation particularly tractable. We note in passing that the time points $\vtime$ are not equally spaced, so that the obvious Fourier transform method for computing the convolution is not available.

The magnitude of $\varp$ is informed by the performance of the pressure gauges, which is well-documented in the literature~\citep{cumming2013assessing}. For a typical well test, the accuracy is estimated to be within $\pm 5$ psi, which we represent by $\sdp \sim \U(0,5)$. For the generation of synthetic data, we take $\sdp = 5$. 

We now consider $P(\vq \given \vtrueq, \know)$. Again, we introduce a variance parameter, and use a multivariate Gaussian model:
\begin{equation}\label{eq:rate_match}
    \vq \given \vtrueq, \varq, \know
    \sim
    \N(\vtrueq, I_{\varq})
    \eqstop
\end{equation}

\noi Expert judgement suggests an associated uncertainty in measured rates of up to $10\%$ of their magnitude. We keep the value fixed to a constant $\sdq = 0.05 \q_m$, where $\q_m=\max_{j}(\{\q_{j}\})$.

Turning to $P(\po \given \truepo, \know)$, we again introduce a variance parameter $\varpo$ and a Gaussian model:
\begin{equation}\label{eq:inpres_match}
    \po \given \truepo, \varpo, \know 
    \sim
    \N(\truepo, \varpo),
\end{equation}

\noi The standard deviation is usually fixed to $\sdpo = 10$, again informed by expert judgement \citep{cumming2013assessing}. 

\subsection{The prior}\label{subs:prior}

We now turn to the distributions $P(\pars \given \know)$, $P(\vtrueq \given \know)$, and $P(\truepo \given \know)$. The independence between these quantities arises from their distinct physical meanings and the available prior knowledge. 

\subsubsection{Prior for model parameters $\pars$}

Absent other information, we assume independence for each of the reservoir and system parameters in $\pars$, producing a distribution of the following form:
    \begin{equation*}
        P(\pars \given \know)= P(P \given \know)\isp P(T \given \know)\isp P(W \given \know)\isp \prod_{i=1}^{n-1} P(R_{i} \given \know)\isp P(M_i \given \know)\isp P(\eta_i \given \know)
    \end{equation*}

\noi for a multi-region radial composite model with $\nregion$ regions. For each component of $\pars$, we approach prior specification from the perspective of eliminating non-physical and implausible values as robustly as possible. Where possible, choices for priors are derived from information in geological and geophysical studies of the corresponding parameters, supplemented by expert knowledge from petroleum engineers. Alternatively, when the parameters of our model do not equate to quantities of particular geological interest, we rely instead on a synthesis of expert judgement, inspection of results from the analysis of other well tests in the literature, and knowledge of the sensitivity of the response function to the parameters as seen in \figref{fig:rcm_params}. Based on these considerations, we use Gaussian priors $T \sim \N(2,0.2^2)$ and $P \sim \N(1.5,0.2^2)$, thereby spanning around two orders of magnitude in the (non-logarithmic) time and pressure scales; while for $W$, we use $W \sim \Ga(1, 0.2)$, thereby spanning many orders of magnitude in the wellbore storage coefficient. 

The $R_i$ represent positive increments in distance from the previous boundary. In order to replicate our prior expectation of homogeneous behaviour in the early stages of the test, but also the assumption that there is reasonable distance between the different transitions, we use $R_i \sim \N(2, 1)$. This means that the expectation of the width of the $i^{\text{th}}$ region is  $100$ times the `effective wellbore radius' $r_{w}e^{-S}$, while the variance means that the width varies between $0.1$ and $10^{5}$ times the effective wellbore radius (up to three standard deviations).

For $\{M_{i}\}$ and $\{\eta_{i}\}$, we adopt a prior mean value of $0$, corresponding to no change between regions, and a symmetric distribution, representing an absence of prior knowledge about whether these quantities will increase or decrease at a transition. To avoid degenerate solutions in which the ratio values grow exceptionally large, we restrict ourselves to a relatively modest variance, and use $M_{i} \sim \N(0, 1)$ and $\eta_{i} \sim \N(0, 1)$, corresponding roughly to a factor of $10^{3}$ variation in each direction.   

\subsubsection{Prior for data parameters}

The individual true rate values are independent $P(\vtrueq \given \know) = \prod_{j=1}^N P(\trueq_j \given \know)$. We use uniform improper priors on $\mathbb{R}_{\geq 0}$ for the components: negative rates represent an injection of fluid into the well, which does not occur in our data. Taking the sign of the rates into account is important, since it might result to changes in the shape of the posterior response function. We use the same prior for $\truepo$.

\subsection{Posterior sampling}\label{sec:sampling scheme}

There is no closed-form expression for $P(\pars, \trueind, \usigma \given \vp, \vq, \po, \know)$, where $\trueind = (\vtrueq, \truepo)$ as before, and $\usigma = (\sdp, \sdq, \sdpo)$, due to the complex form of the response as a function of $\pars$. Consequently, a numerical method such as MCMC is needed to extract information from the posterior.

We can simplify things somewhat, however, by factorizing the posterior: $P(\pars, \trueind, \usigma \given \vx) = P(\trueind \given \pars, \usigma, \vx) \isp P(\pars, \usigma \given \vx)$. We find a Gaussian conditional posterior for $\trueind$:
    \begin{equation}
        P(\trueind \given \pars, \usigma, \vx) \sim \N(A^{-1}b, A^{-1})
        \eqcomma
    \end{equation}

\noi where $A$ and $b$ are given by
    \begin{align*}
        A &= \begin{bmatrix}
           m\precisp + \precispo &
           -1_m^T I_{\precisp} \cmat(\pars)\\
           -\cmat(\pars)^T I_{\precisq} 1_m &
           I_{\precisq} + \cmat(\pars)^T I_{\precisp} \cmat(\pars)
         \end{bmatrix},\\
    \intertext{and}
        b &= \left(\sum_{i=1}^m \vp\precisp + \po\precispo, \vq^T I_{\precisq} -\vp^T I_{\precisp} \cmat(\pars) \right)^T
        \eqcomma
    \end{align*}

\noi where $1_a$ is the $a$-dimensional vector with all elements equal to $1$. The marginal posterior for $\pars$ and $\usigma$ is:
    \begin{multline*}
        P(\pars, \usigma \given \vx) 
        \propto |2\pi A|^{ \frac{1}{2}} 
        \exp\biggl\{
        - m \ln{(\varp)} - N \ln{(\varq)} 
        - \ln{(\varpo)}
        \\
        - \frac{1}{2} \left(
           \vp^T I_{\precisp} \vp +
            \vq^T I_{\precisq} \vq + \po^2\precispo \right) 
        + \operatorname{tr}(b^T A^{-1} b )
        \biggr\}
        \eqstop
    \end{multline*}

\noi Using this factorization, sampling from the posterior can be performed by sampling from the marginal posterior for $\pars$ and $\usigma$, and then sampling from the Gaussian for $\trueind$. This has two advantages: first, that the dimensionality of the space for which MCMC is required is greatly reduced (in our examples, from around $200$ to around $10$); second, that this dimensionality no longer depends on the length of the time series of measurements, which is critical for the scalability of the method. Even for modest sample sizes, the result is a significant reduction in the time needed to generate a given number of samples of $(\pars, \usigma, \trueind)$; in the case that one is interested only in $\pars$, the reduction is even more significant.

Sampling from $P(\pars, \usigma \given \vx)$ requires a careful choice of MCMC algorithm. First, we expect strong correlations between the posterior parameters: Gibbs-type samplers are unsuitable and all parameters need to be updated simultaneously. Second, the covariance structure is not known a priori: the algorithm needs to be adaptive in order to discover the covariance structure as samples accumulate. Finally, we have reason to expect that the posterior will be multimodal, as different combinations of parameter values can yield similar response functions: the algorithm needs to be able to detect the presence of multiple modes and jump between them.

To tackle these challenges, we use the Differential Evolution Markov Chain with snooker updater and fewer chains algorithm (DEzs), as described by \citet{ter2008differential}. The premise is to construct an adaptive random walk Metropolis MCMC using the concept of differential evolution (DE) \citep{storn1997differential}. The MCMC analogue to DE is the Differential Evolution Markov Chain (DEMC) \citep{ter2006markov}, where the candidate solutions in DE correspond to $N$ parallel chains. The proposal for each chain is derived from the remaining $N-1$ chains, which generate the transition probability in each step by updating scale and orientation and, therefore, the covariance structure. DEzs is an extension of DEMC, based on two changes. The first is applied in a randomly selected $90\%$ of the iterations of the chain: in these iterations, the difference vectors may be sampled from the past of the chains, not just the present. The second is applied in the remaining $10\%$ of iterations: the proposal is constructed using a `snooker' update. The advantage of the snooker update is that it jumps relatively easily between modes, thus facilitating sampling from a multi-modal posterior. We use an implementation of the DEzs algorithm from the R package Bayesian Tools \citep{BayesianTools}.




\section{Example: Synthetic channel}
\label{s:SynthChannel}

\begin{figure}
    \centering
        \begin{subfigure}[]{0.49\textwidth}
    \includegraphics[width=\textwidth]{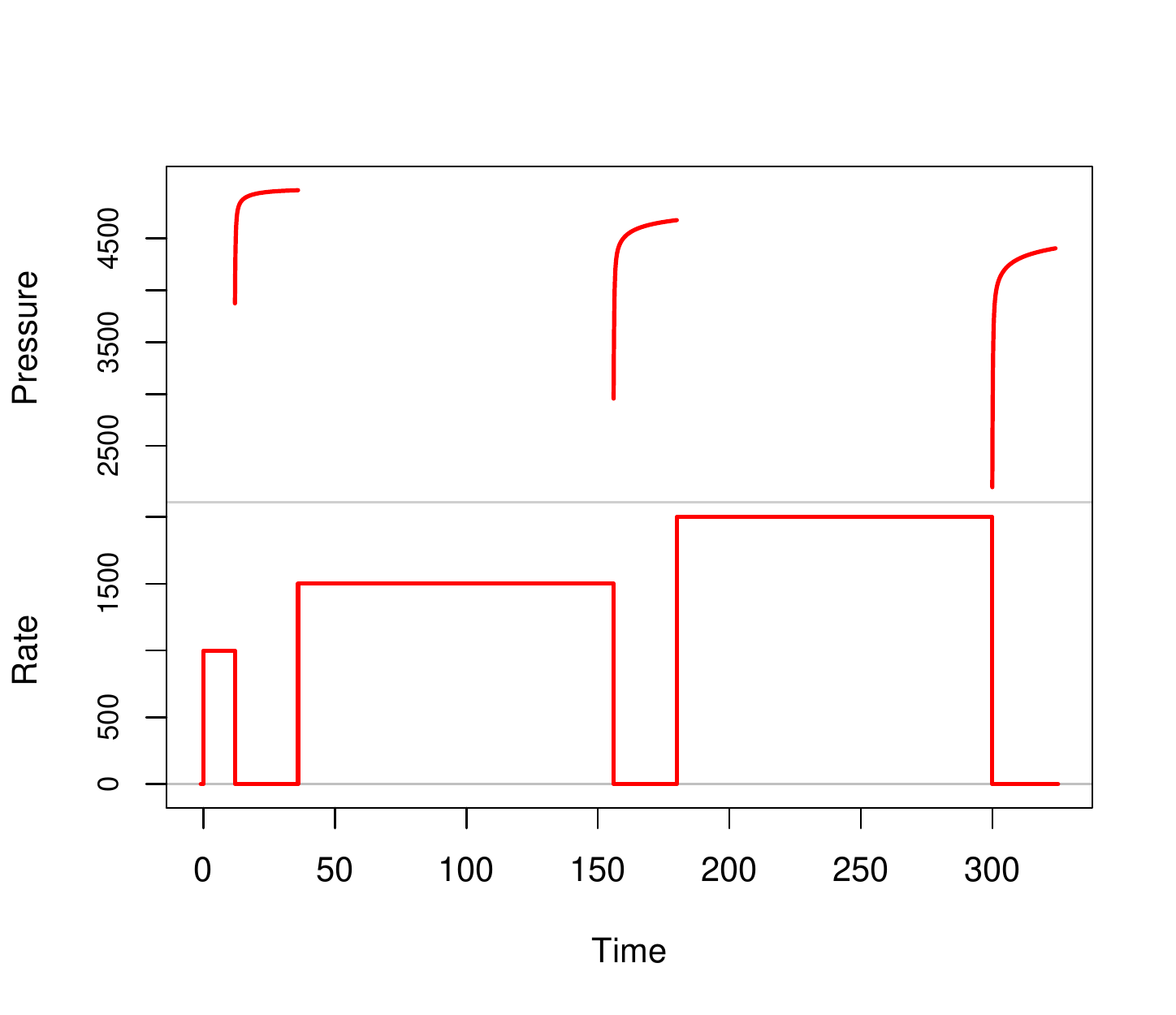}
    \caption{Rate and pressure measurements.}
    \label{fig:channel_data}
        \end{subfigure}
        \begin{subfigure}[]{0.49\textwidth}
    \includegraphics[width=\textwidth]{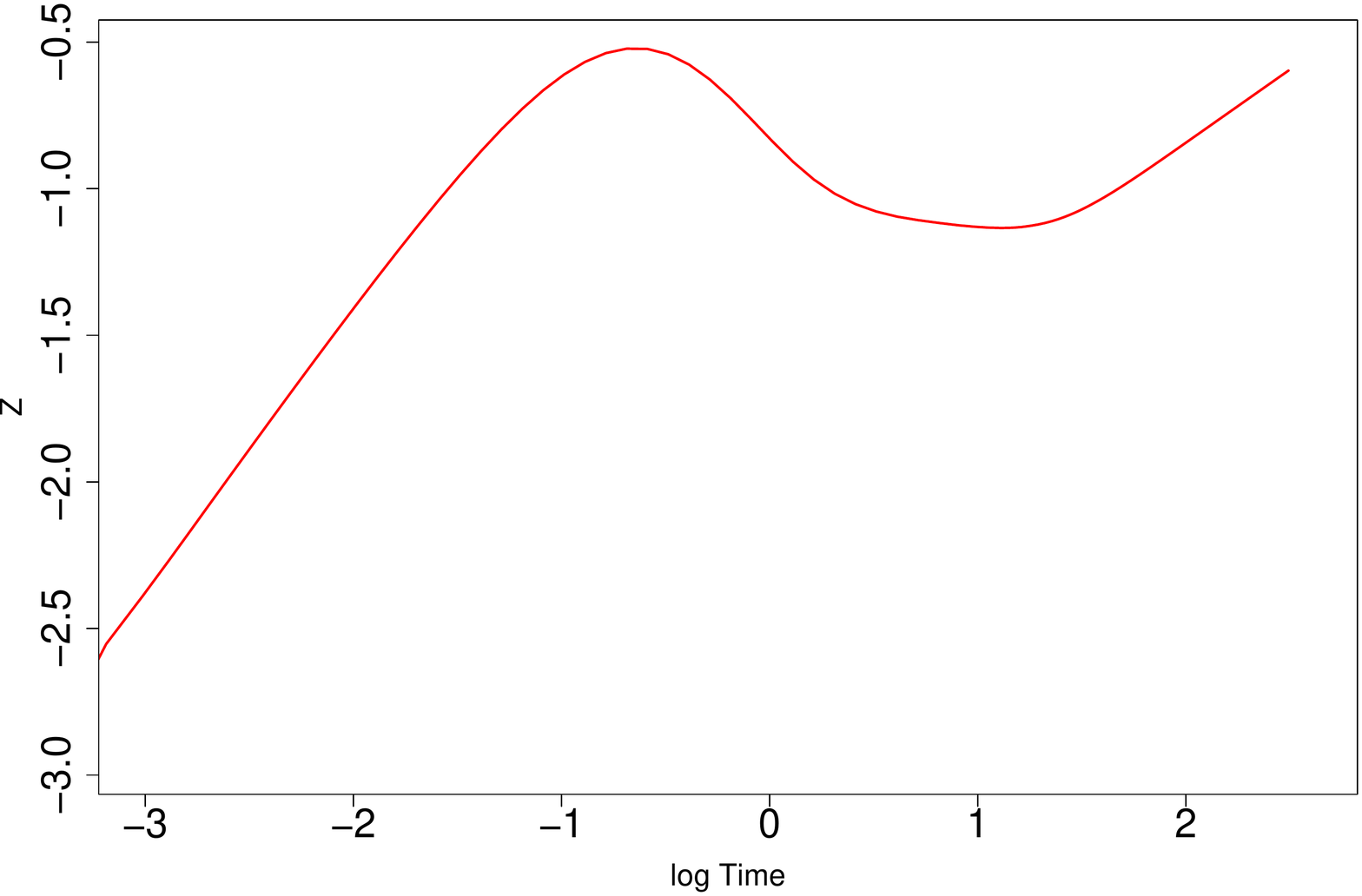}
    \caption{True response.}
    \label{fig:channel_trueresp}
        \end{subfigure}
    \caption{Synthetic channel data.}
    \label{fig:channel_alldata}
\end{figure}

In this section, we validate the method by application to a synthetic reservoir with a known response function with channel behaviour (\figref{fig:channel_trueresp}). We generate a data set (\figref{fig:channel_data}) corresponding to a well test of length 320 hours, comprising three periods of production at constant but different rate values, and a total of 272 pressure measurements generated by convolution of the rate with the known response. To represent potential observational error, the pressure measurements include additive random noise with known variance $\varp = 25$.

Our primary focus with this analysis is verify that our methods are able to capture the reservoir response behaviour, even though a channel is not a radially composite reservoir, and to determine how precisely the parameter values can be obtained from the data given only weakly informative priors. Consequently, we treat the true rate values and initial pressure as known, and equal to their measured values, leaving $\pars$ as the parameter of interest, and we choose vaguer priors for $\pars$ than those recommended in \secref{subs:prior}: a Gamma prior for $W$ and broad uniform priors for the remaining parameters. The DEzs algorithm was run with three parallel chains for 666,667 iterations. A thinning of 50 and a burn-in of 1,000 were applied, leaving 37,002 iterations to be used for inference.
\begin{figure}[t]
    \centering
        \begin{subfigure}[]{0.49\textwidth}        
            \includegraphics[width=\textwidth]{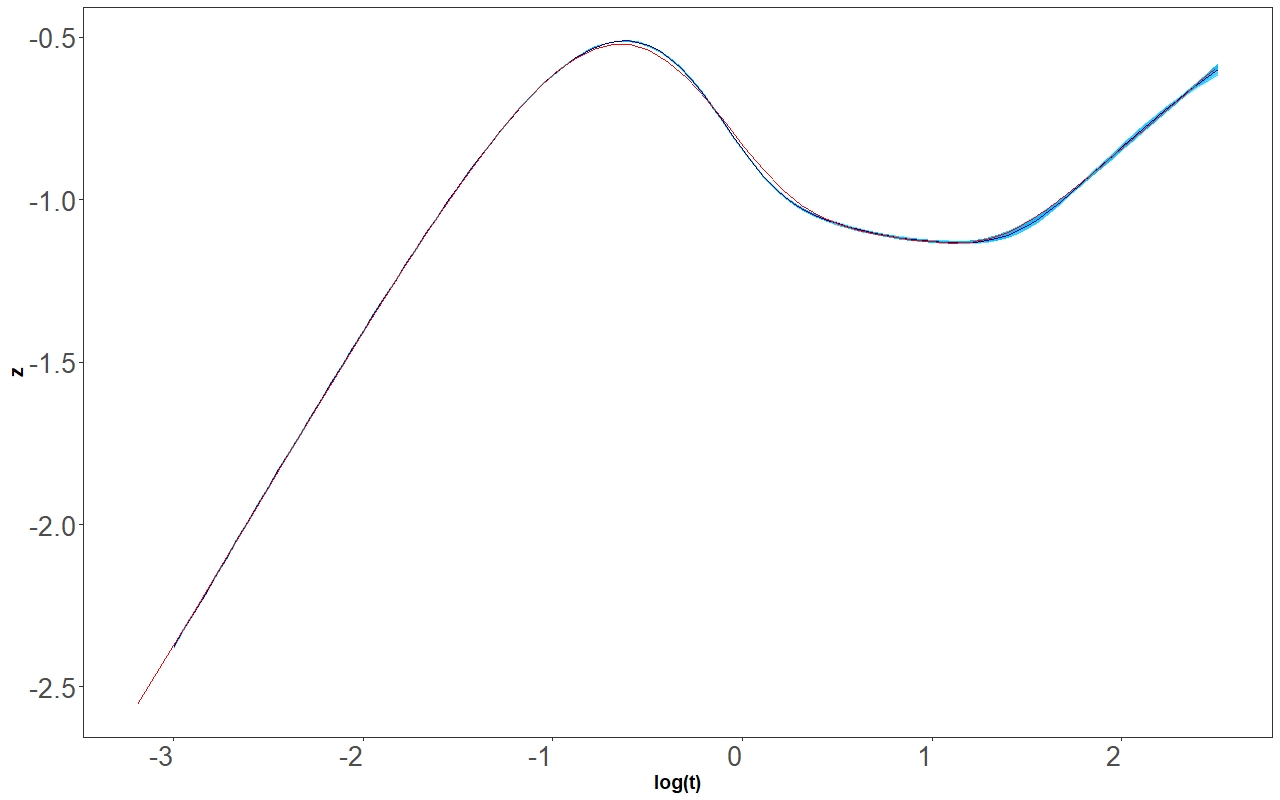}
            \caption{Posterior response functions.}
            \label{fig:channel_resp}
        \end{subfigure}
        \begin{subfigure}[]{0.49\textwidth}        
            \includegraphics[width=\textwidth]{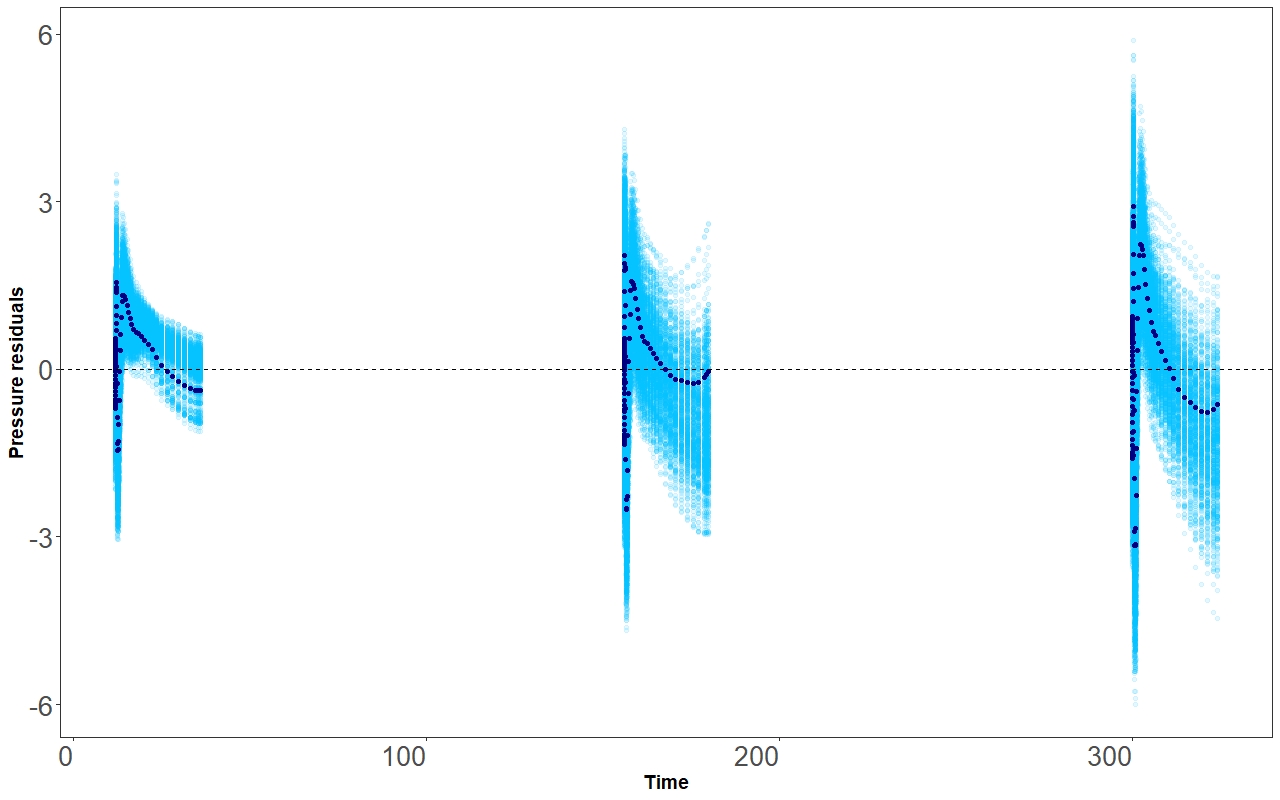}
            \caption{Posterior pressure residuals.}
            \label{fig:channel_resids}
        \end{subfigure}
        \begin{subfigure}[]{0.49\textwidth}
            \includegraphics[width=\textwidth]{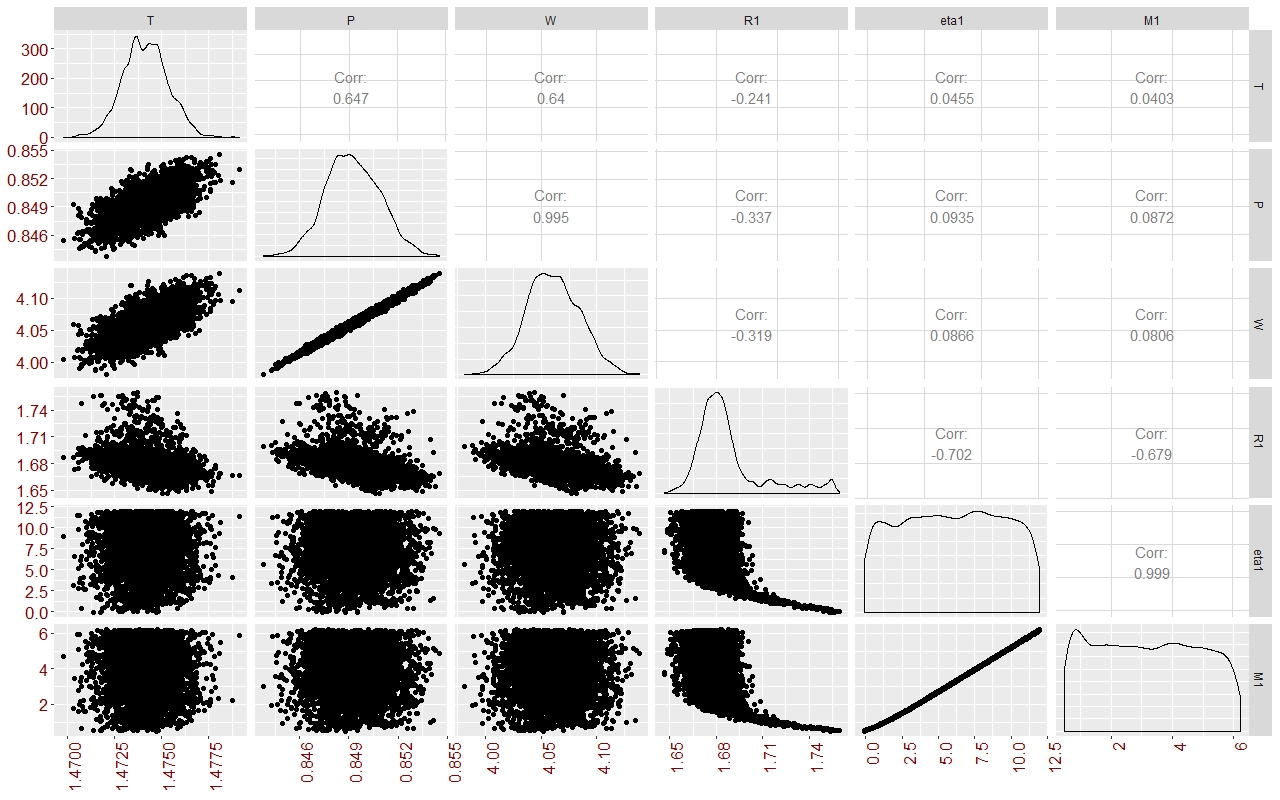}
            \caption{Scatterplots.}
            \label{fig:channel_scatter}
        \end{subfigure}
        \begin{subfigure}[]{0.49\textwidth}
        \includegraphics[width=\textwidth]{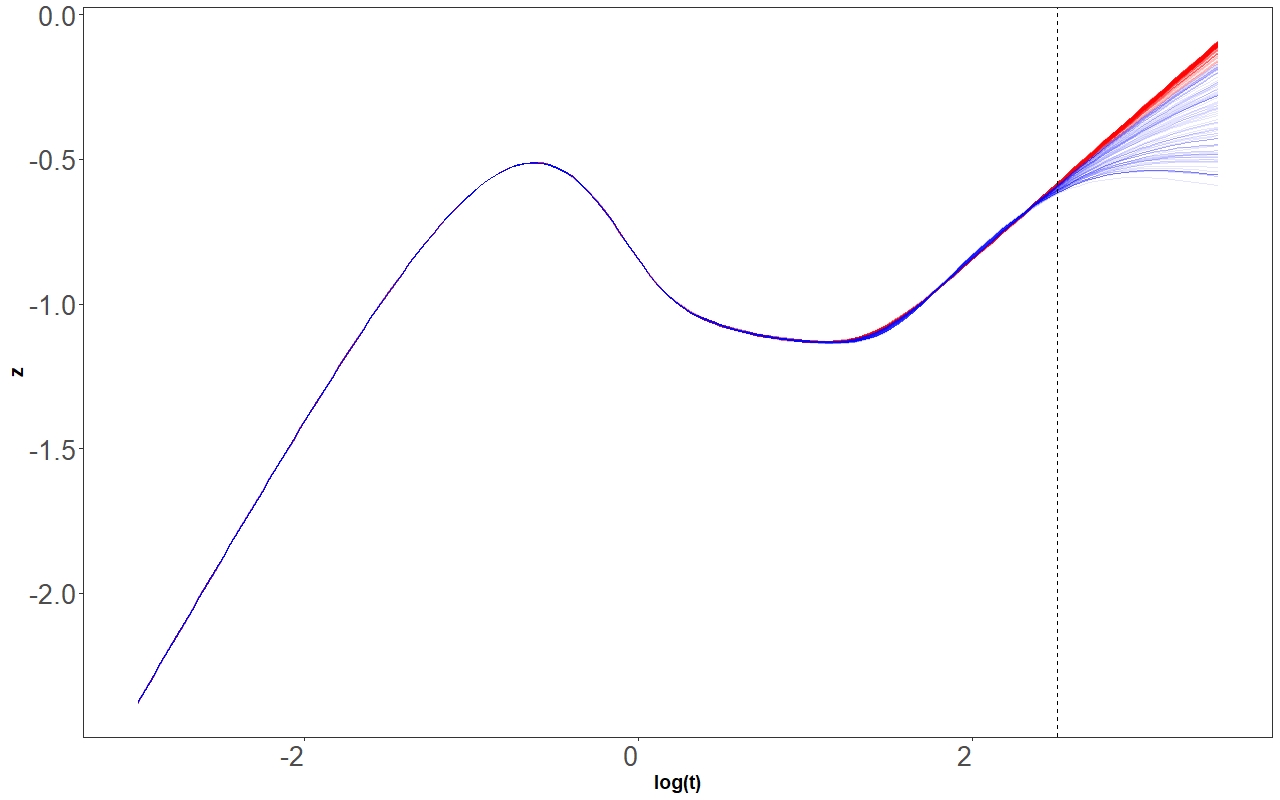}
        \caption{Extended response functions.}
        \label{fig:channel_extension}
        \end{subfigure}
    \caption{Posterior marginal distributions, posterior response functions, pressure residuals, and extended response functions, for the posterior response parameters in the synthetic channel model.}
    \label{fig:channel_output}
\end{figure}

The resulting response function posterior samples are shown in \figref{fig:channel_resp}, as light blue curves; the MAP response is shown in dark blue and the true channel response is overlaid in red. All samples show the characteristic limiting half-unit slope expected of a channel reservoir, thereby validating the response shape. \figref{fig:channel_resids} shows the residuals between the posterior samples of reservoir pressure and the true values from the underlying synthetic model. Most of the samples lie within $\pm 3$ psi of the true pressures, which is well within usual tolerances for such analyses.

\figref{fig:channel_scatter} shows summaries of the posterior MCMC samples of the response parameters giving rise to the response curves in \figref{fig:channel_resp}. The marginal posteriors for $T$, $P$, $W$, and $R$ have substantially smaller variances than their priors, indicating that the data identifies these quantities quite precisely. In contrast, the marginal posteriors for $M$ and $\eta$ are generally close to their prior forms, indicating an inability to identify specific values for these transition parameters. This picture is refined by the joint posterior marginals, shown as scatterplots in \figref{fig:channel_scatter}. First, we see that there is a strong positive correlation between $P$ and $W$, and weaker positive correlations between both those parameters and $T$. These can be explained by the fact that small changes in the early-time shape ($W$) can be replicated by small shifts in the response ($T$ and $P$), \eg larger $P$ values shift the response curve downwards, while larger $W$ values make the `bump' steeper; these effects cancel each other over a narrow range.

The late-time parameters $(R, M, \eta)$ also show interesting behaviour. The $(M, \eta)$ plot shows that, unlike $T$, $P$, $W$, and $R$, these parameters cannot be well identified individually, \ie cannot be constrained to a quasi-$0$-dimensional subset, but they can be constrained to a quasi-$1$-dimensional subset, and in fact, the relationship is essentially affine. However, there are two separate regimes. For smaller $R$, where the bulk of the probability lies, $M$ and $\eta$ can take on a large range of values (although themselves still linearly related); for larger $R$, $M$ and $\eta$ are essentially determined as a function of $R$. There is a clear physical explanation for these two regimes, best explained by reference to \figref{fig:channel_extension}. This shows the posterior response function samples from \figref{fig:channel_resp} plotted over an extended range; the red curves correspond to the first regime, the blue to the second. In the first, the limiting slope is indeed present. Its time of onset (essentially $R$) is quite well specified, but provided the change in rock properties passes some threshold, the exact values of $M$ and $\eta$ do not matter. In the second, the limiting slope is an illusion produced by the window of observation; in reality, there is a jump in the response function to a constant higher level. In this regime, the time of onset and the values or $M$ and $\eta$ are tightly coupled: the smaller the jump, the later must be the onset to push it out of the observation window. These results reveal that the standard industry interpretation of the half-unit slope as exclusively indicative of a channel is too definitive: a range of other reservoir configurations are consistent with this behaviour.

\section{Example: Field data from an oil reservoir}
\label{sec:oil_data}

\begin{figure}[t]
\centering
    \includegraphics[width=.5\textwidth]{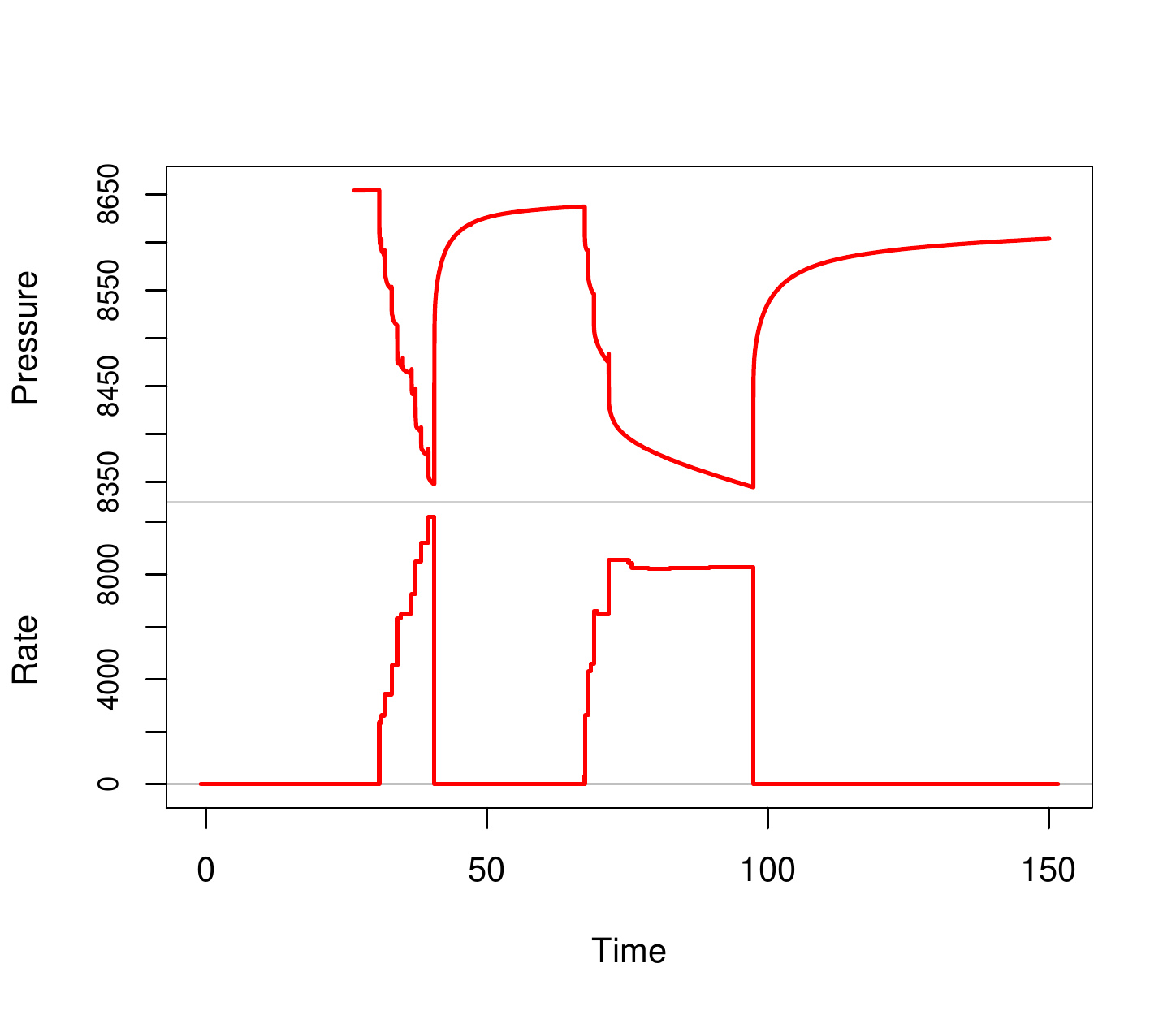}
    \caption{Field oil data rate and pressure measurements}\label{fig:oil_data}
\end{figure}

In this section, we analyse the results of a real oil field dataset of 2,273 pressure observations and 22 production rates over approximately 150 hours, shown in \figref{fig:oil_data}. We consider radial composite models with 1--4 transitions (2--5 regions). The true rates and initial pressure are unknown, so that in addition to using the DEzs algorithm for $\pars$ and $\sdp$, we sample from the Gaussian for $\vtrueq$ and $\truepo$, as described in \secref{sec:sampling scheme}. We ran three parallel chains in the DEzs algorithm for 1,666,667 iterations each, which, after thinning of 50 and burn-in of 10,000 were applied, resulted in 70,002 samples for inference.

\begin{figure}[t]
    \centering
        \begin{subfigure}[]{0.49\textwidth}
        \includegraphics[width=\textwidth]{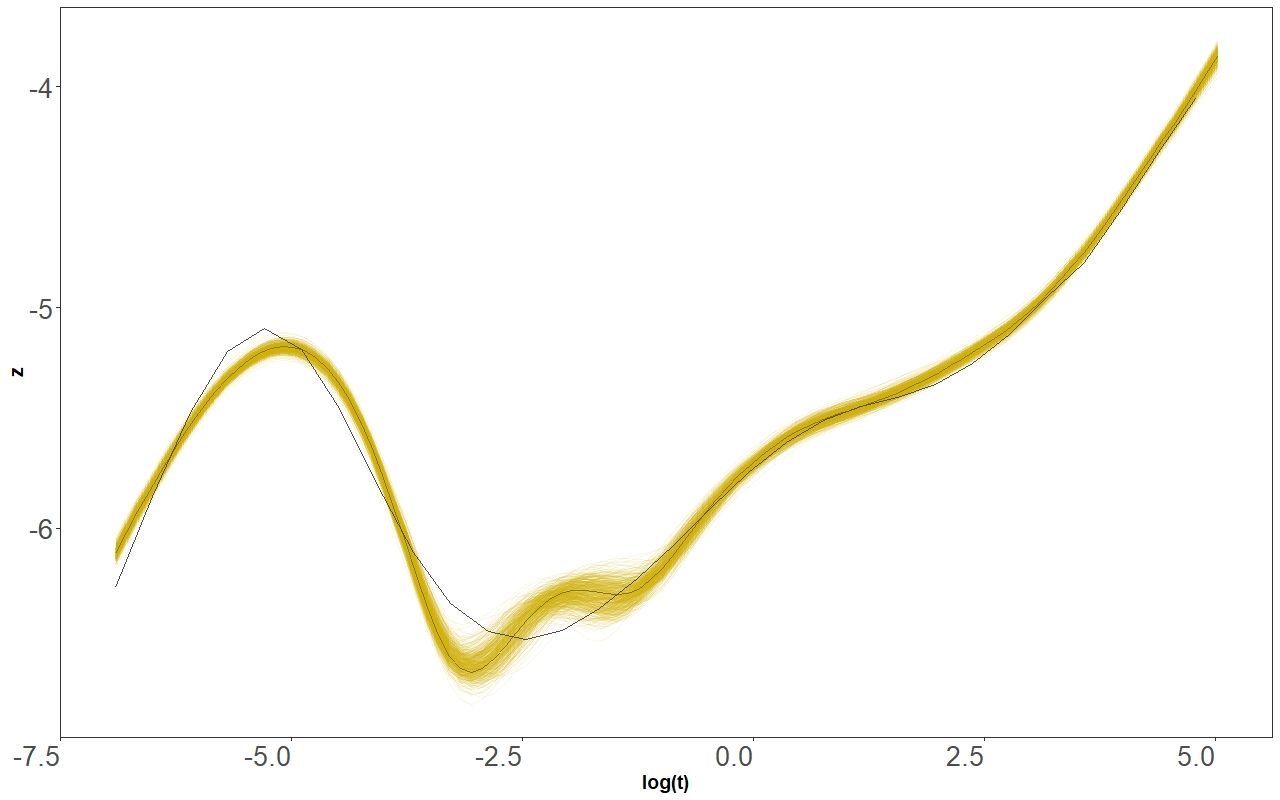}
        \caption{Posterior response functions.}
        \label{fig:posterior_responses}
        \end{subfigure}
        \begin{subfigure}[]{0.49\textwidth}        
        \includegraphics[width=\textwidth]{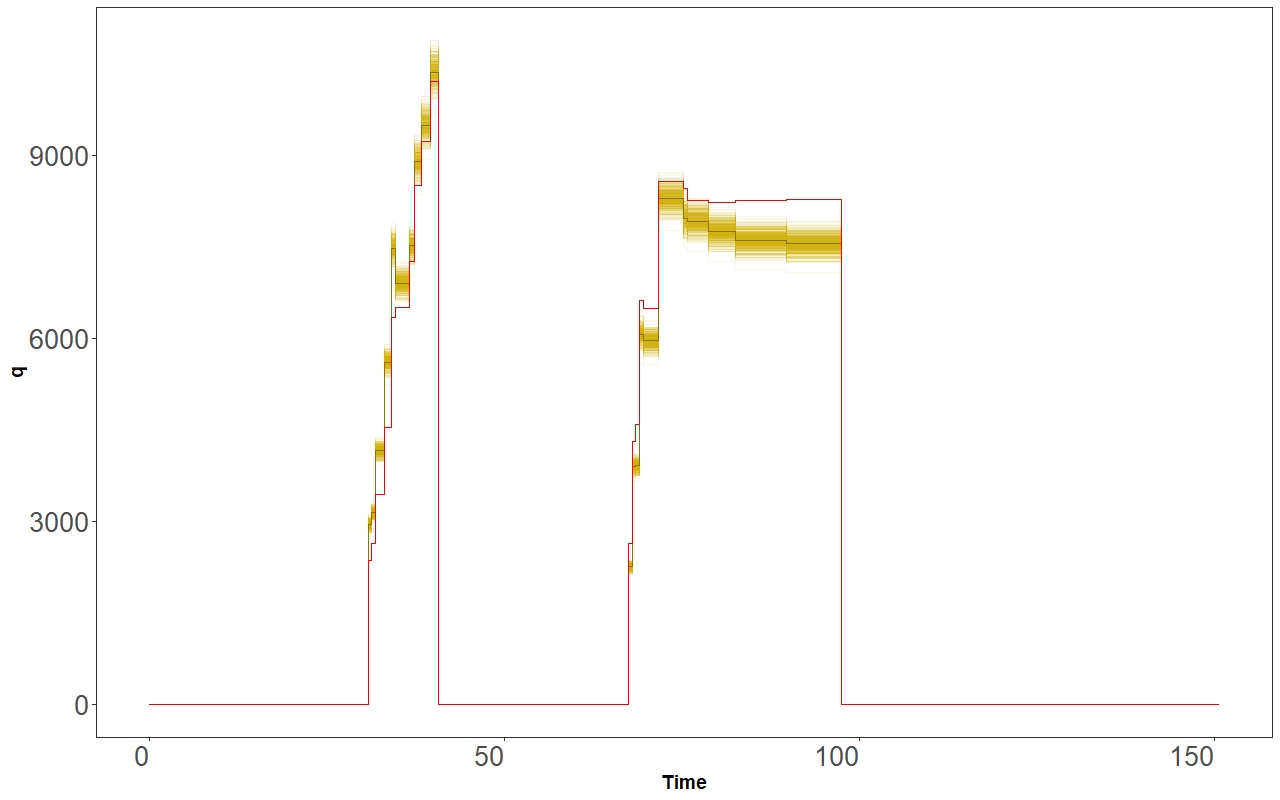}
        \caption{Posterior rates.}
        \label{fig:posterior_rates}
        \end{subfigure}

    \caption{Posterior response and rate samples for the 3-transition model.}
    \label{fig:3region_output}
\end{figure}

For the following discussion, we focus on the 3-transition model. The posterior response and rate samples are shown in \figref{fig:3region_output}. The response shows a very early feature, at around $\logt = -2.5$, suggesting that the wellbore is very close to an inhomogeneity. The rate samples in \figref{fig:posterior_rates} show significant deviation from the measured values, although the general structure remains. This is consistent with the known scale of errors in these measurements. 

\begin{figure}[t]
    \centering
        \includegraphics[width=\textwidth]{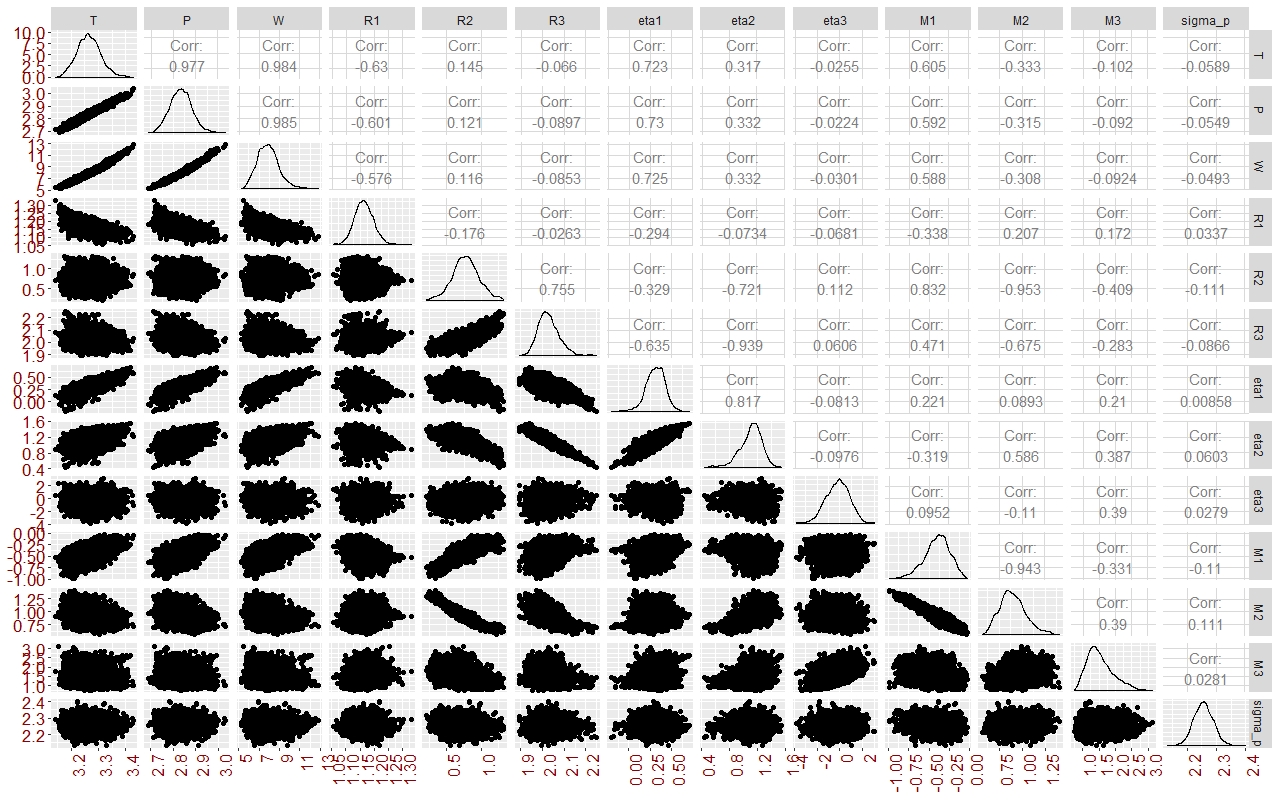}
        \caption{Scatterplots for the response parameters.}
  \label{fig:oil_all}
\end{figure}

The one- and two-dimensional marginals are shown in in \figref{fig:oil_all} as smooth density estimates and scatterplots respectively. The marginals for $T$, $P$, and $W$ tend to high values relative to their priors, but although their posterior variances are larger than in the synthetic case, as one might expect, they are still small relative to the priors. The radii parameters are reasonably far apart, suggesting that there is no degeneracy arising from a trivial region. The $\truepo$ marginal suggests that the true initial pressure is larger than the corresponding measurement, but the difference, around $2\text{psi}$, is not unreasonable. The variance $\sdp$ lies in the range $[2, 2.5]$, which is certainly a reasonable pressure measurement error for an oil field reservoir. The posterior variances of the late-time posterior parameters are small, with the exception of the ratio parameters for the third transition, which cover a significant portion of their prior ranges. This is behaviour analogous to that found in the synthetic case; we discuss it further presently.

The scatterplots show a more complicated correlation structure than the synthetic results. The strong positive correlation for the early-time parameters extends to $T$, which implies that the horizontal shift of the response has a more significant role in compensating for the effects of $P$ and $W$. Specifically, larger values of $T$ shift the curve to the left, causing the skin `bump' to appear earlier, which, in consequence, causes $W$ to increase. 

There are also numerous late-time parameter correlations, some of which are reasonably interpretable. For example, we detect negative correlations between $M_2$ and the parameters $M_1$ and $R_{2}$, indicating that larger first transitions correspond to smaller second transitions. In terms of the response, this says that the earlier the effect of the second transition arrives, the larger the size of the transition needs to be to produce similar results.

\subsection{Additional visualizations}

\begin{figure}[t]
    \begin{subfigure}[]{0.49\columnwidth}   
        \includegraphics[width=\textwidth]{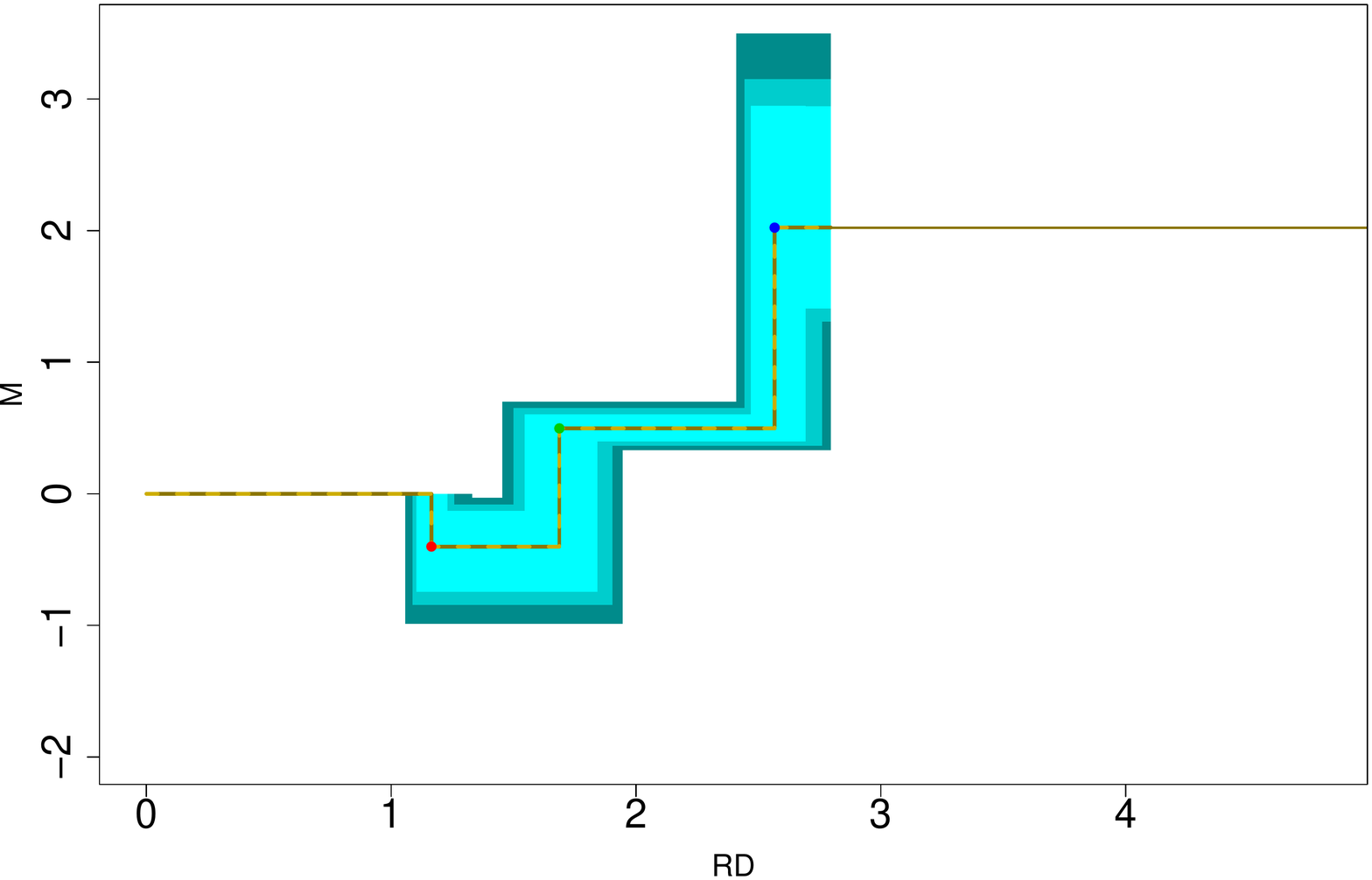}
        \caption{Uncertainty step plot for $m$.}
        \label{fig:3r_stepplots_M}
    \end{subfigure}
    \begin{subfigure}[]{0.49\columnwidth}   
        \includegraphics[width=\textwidth]{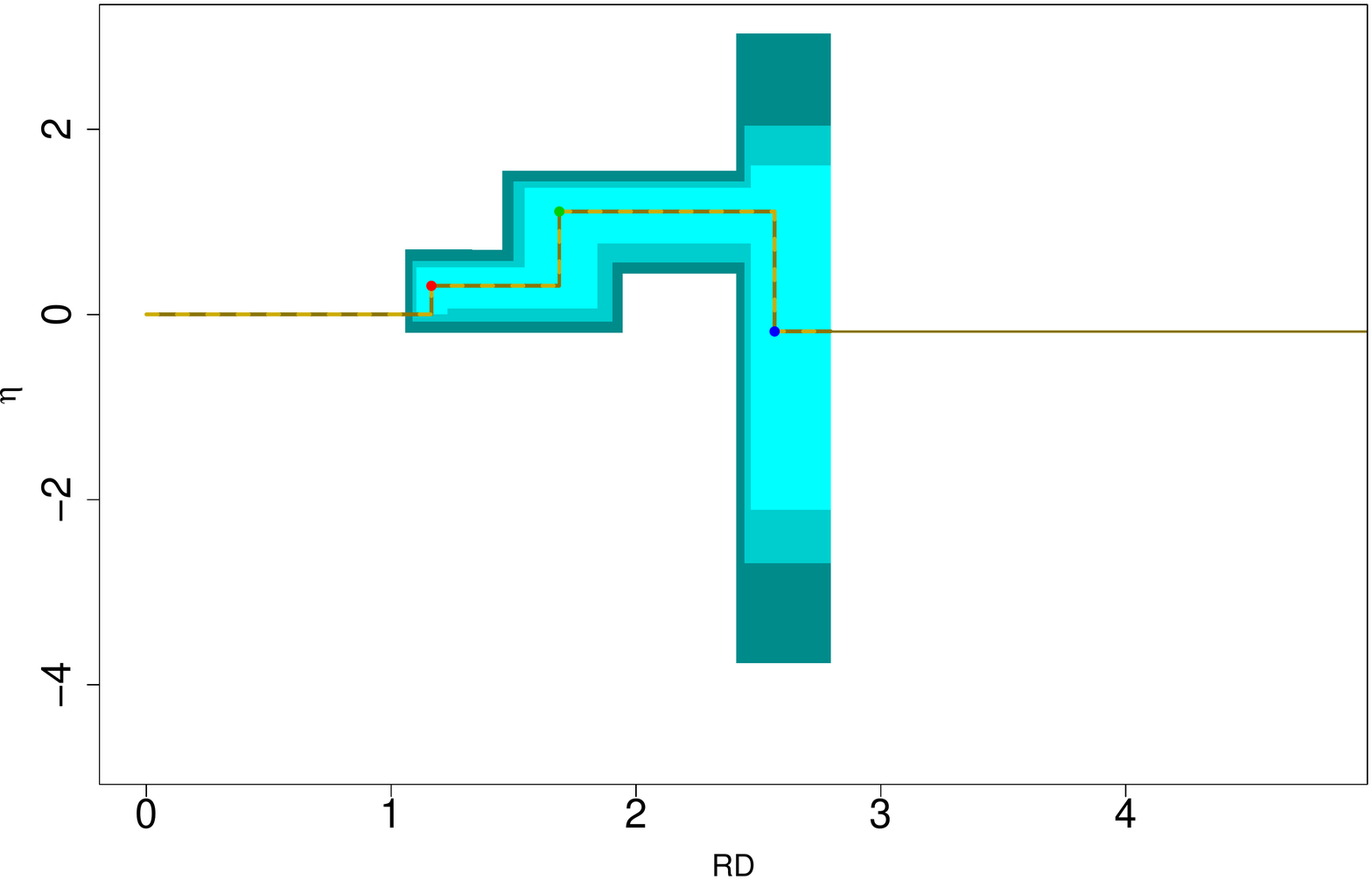}
        \caption{Uncertainty step plot for $\eta$.}
        \label{fig:3r_stepplots_eta}
    \end{subfigure}
    \caption{Uncertainty step-plots for the 3 transition model applied to the oil field dataset.}\label{fig:oil_step}
\end{figure}

\figref{fig:oil_step} shows an alternative representation of the posterior for the reservoir parameters, showing how mobility and diffusivity change with radial distance from the wellbore. Specifically, defining $\rho_{i} = \log\sum_{j = 1}^{i} e^{R_{j}} = \log\bigl({r_{i} \over r_{w}e^{-S}}\bigr)$ and $m_{i} = \sum_{j= 1}^{i-1} M_{j} = \log\bigl({(k/\mu)_{1} \over (k/\mu)_{i}}\bigr)$, we plot the piecewise constant functions taking $[\rho_{i-1}, \rho_{i})$ to $m_{i}$ and $\eta_{i-1}$ respectively. In these plots, a solid line indicates the posterior mean, while the coloured bands depict the posterior sample range (dark blue), $99\%$ credible posterior intervals (middle blue), and $95\%$ credible posterior intervals. In general terms, the step changes in these parameters correspond to the features in the response function after the early-time skin effect. 

The first change in parameter values consists of a small decrease in $\eta$ and an increase in $m$: this produces the increase in the response curve at around $\tau=-2$ in \figref{fig:posterior_rates}. The subsequent increase in the response function corresponds to the second, larger change in $m$ and the switch to positive values for $\eta$. Finally, the late-time unit slope observed in the response function corresponds to the third, more substantial positive shift in $m$, and the wide range of possible values of $\eta$. These changes are indicative of an impermeable boundary to the reservoir, and this is indeed the standard interpretation of a late-time unit slope. 

\subsection{Model comparison}

\begin{figure}[t]
\centering
    \begin{subfigure}[]{0.49\columnwidth} \includegraphics[width=0.99\textwidth]{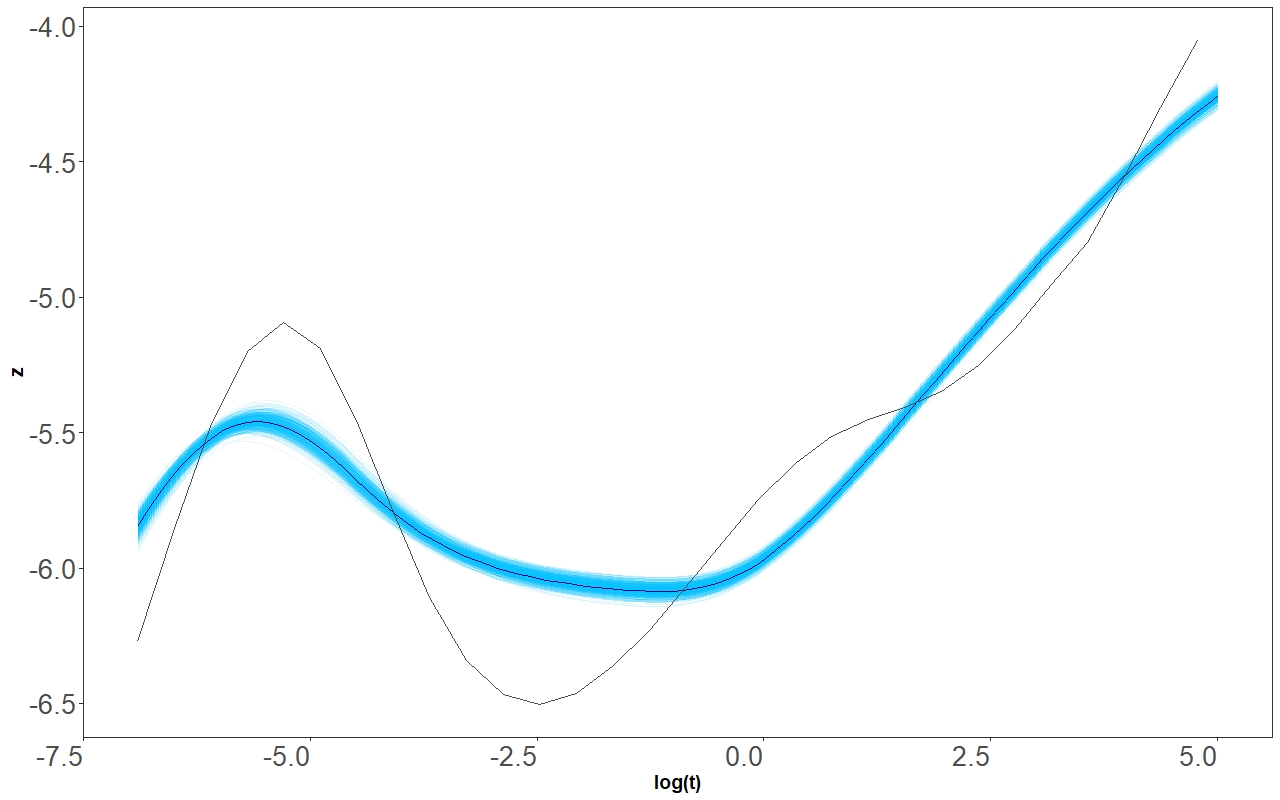}
    \caption{1 transition model.}
    \end{subfigure}
    \begin{subfigure}[]{0.49\columnwidth} 
    \includegraphics[width=0.99\textwidth]{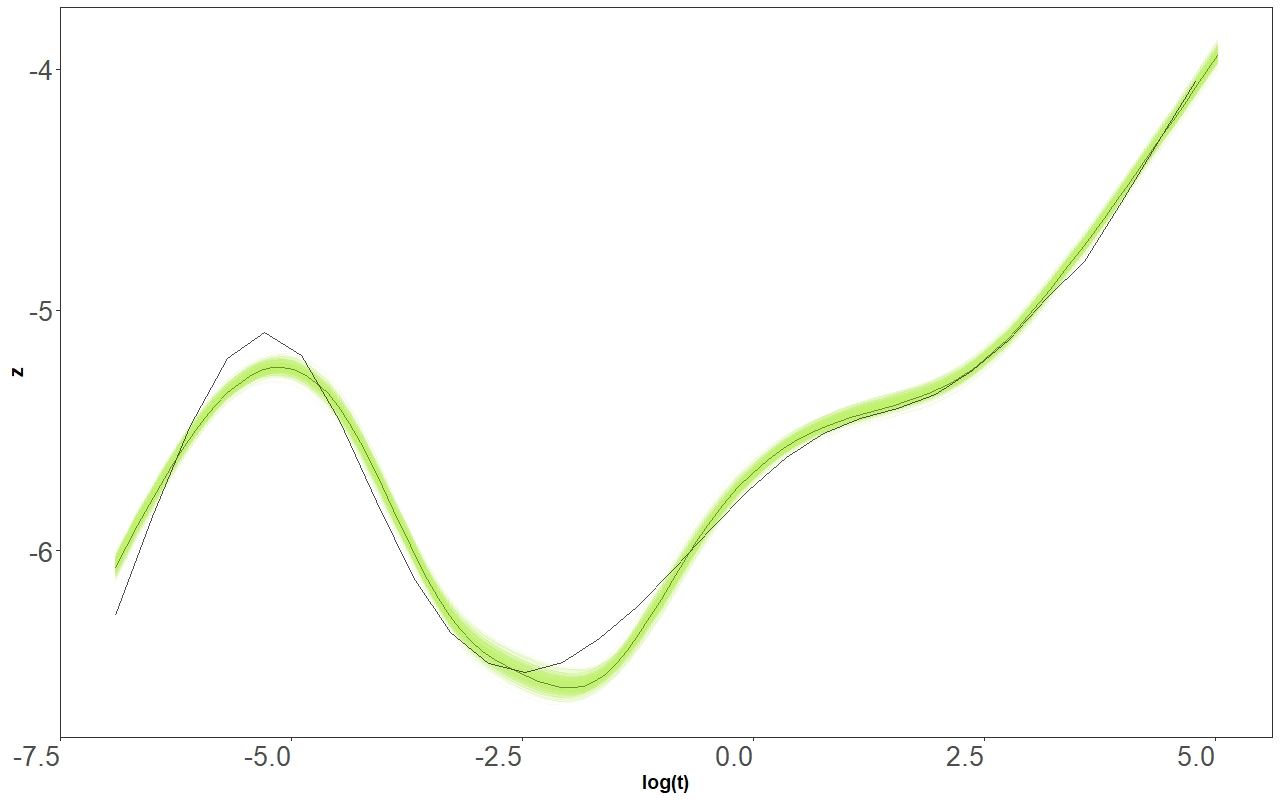}
        \caption{2 transition model.}
    \end{subfigure}
    \begin{subfigure}[]{0.49\columnwidth} 
\includegraphics[width=0.99\textwidth]{3r_respf.jpeg}
    \caption{3 transition model.}
    \end{subfigure}
    \begin{subfigure}[]{0.49\columnwidth} 
    \includegraphics[width=0.99\textwidth]{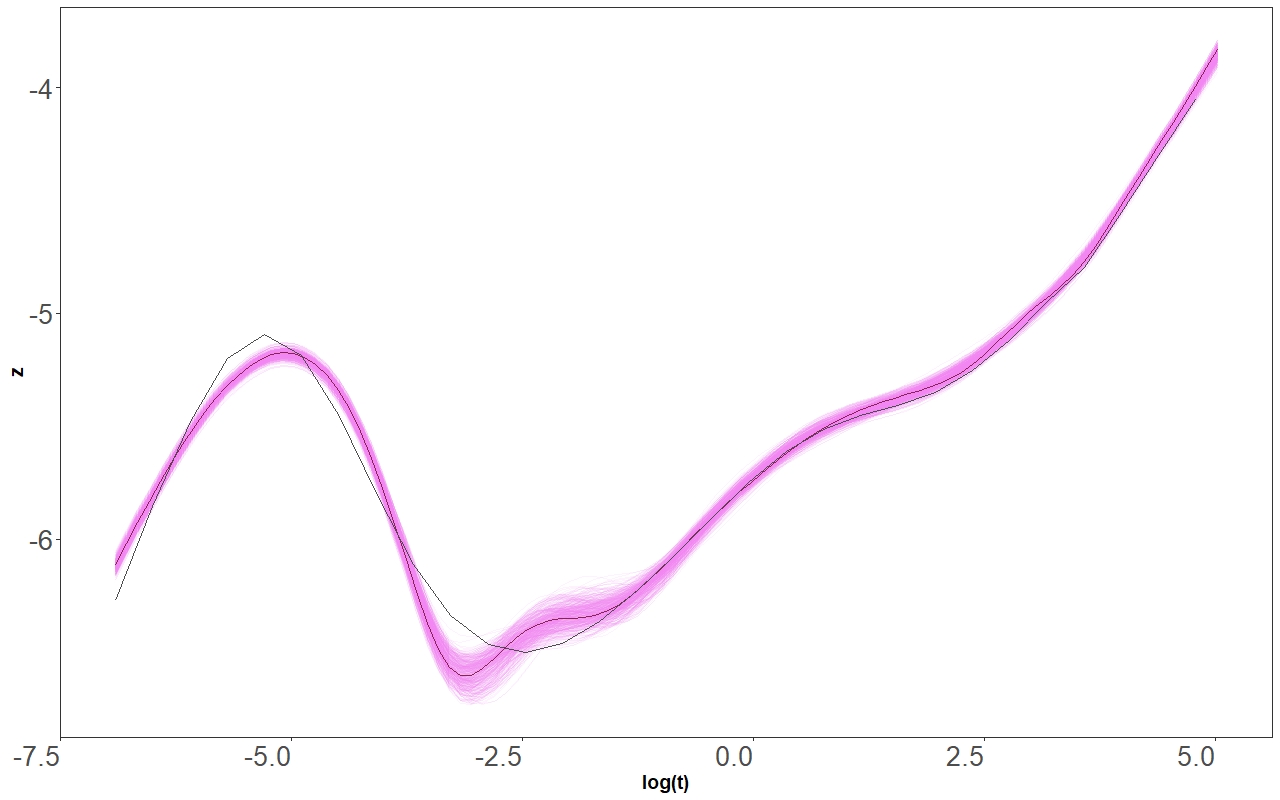}
        \caption{4 transition model.}
    \end{subfigure}
\caption{Response uncertainty for 1, 2, 3 and 4 transitions respectively including the TLS result (as a single black curve).}
\label{fig:3r_resp_uncty}
\end{figure}

In \figref{fig:3r_resp_uncty} we plot the posterior response function samples for models with 1--4 transitions. Each plot includes a black curve showing the TLS result, the current state-of-the-art. The 1-transition result looks somewhat different to and flatter than the other results, probably indicating an under-parameterization giving an over-smoothed result. The 2-transition result is closest to the TLS result, with slight deviations in early times, again probably from the relative rigidity of the model. The 3- and 4-transition results are very similar to each other, suggesting that the 4-transition model is over-parameterized and that the 3-transition model is an adequate description. The 3- and 4-transition results show a qualitatively new feature, occurring after the initial skin `bump', as compared to the 2-transition and the TLS results. This feature could be indicative of hemiradial flow, which occurs when a boundary is detected close to the wellbore. Note that the TLS method has smoothed over this feature.

In terms of computational cost, there is a significant difference between the TLS and our method, with the former taking usually a few minutes and the latter a few hours, depending on the dataset. However, this trade-off comes with some important inferential advantages; namely the uncertainty quantification, the exclusion of non-physical results, and the access to meaningful parameters.

In \tabref{oil:modelselec_freq}, we tabulate the AIC \citep{akaike1998information}, BIC \citep{schwarz1978estimating}, DIC \citep{spiegelhalter2002bayesian}, and marginal likelihood \citep{chib2001marginal} values for all four models. All criteria agree in preferring the 3-transition model, although the 4-transition models is a close second. The results for the 3- and 4-transition results corroborate the idea that the 4-transition model is over-parameterized, and that three transitions are sufficient to explain the data in this case.

It is worth noting that alternative algorithms, such as the reversible jump MCMC in \cite{green1995reversible} can potentially facilitate model selection, by overcoming the requirement of running multiple chains and the subsequent step of using information criteria and Bayes factors.
\begin{table}
\centering
\begin{tabular}{lrrrr}
  \hline
 Criterion & 1 & 2 & 3 & 4 \\ 
  \hline
  AIC & 12781.58 & 10559.51 & \textbf{10454.77} & 10457.10 \\ 
  BIC & 12832.06 & 10631.62 & \textbf{10548.51} & 10572.47 \\ 
  DIC & 12781.36 & 10559.41 & \textbf{10453.53} & 10459.58 \\ 
  Neg log marginal likelihood & 6405.793 & 5299.607 & \textbf{5246.078} & 5249.384 \\
   \hline
\end{tabular}
\caption{Information criteria and marginal likelihoods for results from models with 1--4 transitions. Bold indicates the smallest value for each criterion.}
\label{oil:modelselec_freq}
\end{table}

\section{Concluding remarks}
\label{conclusion}

We have presented a new method for the deconvolution of reservoir well test data, using a parameterized, physical model for the response function, embedded in a fully Bayesian context in order to deal with observational error and other uncertainties. We applied the method to synthetic and field data sets, making inferences about the response function and other quantities of interest. We examined the effect of different numbers of regions on the same data set and the contribution of the parameters to the variation of the system. Finally, we developed new forms of visualisation for the posterior parameters and their uncertainties.

The proposed method has a number of advantages over the current state-of-the-art TLS method: the parameters can be linked to the flow behaviour of the reservoir, and thus can either confirm results found with other well testing methods, or provide further information for expert interpretation; the use of full distributions for the parameters of interest provides a principled way of evaluating joint uncertainty; and the method guarantees physically sensible results.

We note that the methods described here can also be applied to gas reservoir data, after performing a change of units to `pseudo-pressure' in order to take into account the variable compressibility and viscosity of gas, which also produces corresponding changes in the priors for $T$ and $P$.

Next steps include developing faster methods to compute the response and, more substantially, the extension to the multi-well case \citep{cumming2013multiple}.

\section{Acknowledgements}
\label{Acknowledgements}
This work was performed as part of several Imperial College Joint Industrial Projects (JIP) sponsored by Anadarko, BHP Billiton, BP, Chevron, ConocoPhillips DEA Group, Engie, ENI, Inpex, Nexen CNOOC, Occidental Petroleum, Paradigm, Petro SA, Petrobras, Saudi Aramco, Schlumberger, Shell, Weatherford and Total. 

\bibliographystyle{rss}
\bibliography{mybibliography}

\end{document}